\documentclass[10pt,journal,compsoc]{IEEEtran}
\usepackage{amsmath}
\usepackage{amssymb}
\usepackage{cite}
\usepackage{algorithm,algorithmic}
\usepackage{subfigure}
 \usepackage{multirow}
 \usepackage[table,xcdraw]{xcolor}
 \usepackage[normalem]{ulem}
 \useunder{\uline}{\ul}{}
\usepackage[table]{xcolor}
\usepackage{tikz}
\usepackage{url}
\usetikzlibrary{fit,positioning}
\newcolumntype{L}[1]{>{\raggedright\arraybackslash}p{#1}}
\newcolumntype{C}[1]{>{\centering\arraybackslash}p{#1}}
\newcolumntype{R}[1]{>{\raggedleft\arraybackslash}p{#1}}

\ifCLASSOPTIONcompsoc

\DeclareMathOperator*{\minimize}{minimize}
\newcommand\BibTeX{{\rmfamily B\kern-.05em \textsc{i\kern-.025em b}\kern-.08em
T\kern-.1667em\lower.7ex\hbox{E}\kern-.125emX}}
\else
 \usepackage[nocompress]{cite}
\fi
\ifCLASSINFOpdf
\else
\fi
\allowdisplaybreaks
\usepackage{hyperref}
\begin{document}
\title{Joint Radio Resource Allocation and Cooperative Caching in PD-NOMA-Based HetNets}
	\author{Maryam~Moghimi, \IEEEauthorblockN{Abolfazl Zakeri, \IEEEmembership{Member IEEE}, Mohammad Reza Javan, \IEEEmembership{Senior Menmber,
			IEEE}, Nader Mokari, \IEEEmembership{Senior Member, IEEE}, and Derrick~Wing~Kwan~Ng,
		\IEEEmembership{Senior Member, IEEE}}
	\IEEEcompsocitemizethanks{\IEEEcompsocthanksitem M. Moghimi, A. Zakeri, and N. Mokari are with the Department of ECE,
		Tarbiat Modares University, Tehran, Iran (email: \{maryammoghimi, abolfazl.zakeri, and nader.mokari\}@modares.ac.ir).\protect\\
		Mohammad R. Javan is with the Department of Electrical and Robotics Engineering, Shahrood University of Technology,
		Shahrood, Iran (javan@shahroodut.ac.ir).\protect\\
		D. W. K. Ng is with the School of Electrical Engineering and Telecommunications, The University of
		New South Wales, Australia(E-mail:w.k.ng@unsw.edu.au).
	}}
%

\IEEEdisplaynontitleabstractindextext
\IEEEpeerreviewmaketitle
\IEEEtitleabstractindextext{
	\begin{abstract}
In this paper, we propose a novel joint resource allocation and cooperative caching scheme for power-domain non-orthogonal multiple access (PD-NOMA)-based heterogeneous networks (HetNets). In our scheme, the requested content is fetched directly from the edge if it is cached in the storage of one of the base stations (BSs), and otherwise is fetched via the backhaul.
Our scheme consists of two phases: 1. Caching phase where the contents are saved in the storage of the BSs, and
2. Delivery phase where the requested contents are delivered to users. We formulate a novel optimization problem over radio resources and content placement variables. We aim to minimize the network cost subject to quality-of-service (QoS), caching, subcarrier {assignment}, and power allocation constraints. By exploiting advanced optimization methods, such as alternative search method (ASM), Hungarian algorithm, successive convex approximation (SCA), we obtain an efficient sub-optimal solution of the optimization problem. {Numerical results illustrate that our ergodic caching policy via the proposed resource management algorithm can achieve a considerable reduction on the total cost on average	compared
to the most popular caching and random caching policy. Moreover, our cooperative NOMA scheme outperforms orthogonal multiple access (OMA) in terms of the delivery cost in general with an acceptable complexity increase. }
\end{abstract}
\begin{IEEEkeywords}
Cooperative caching, resource allocation, power-domain non-orthogonal multiple access (PD-NOMA), successive convex approximation (SCA).
\end{IEEEkeywords}}
	\vspace{5em}
	\maketitle
	\section{Introduction}
\IEEEPARstart{T}{he} tremendous growth of multimedia applications with high demand in data rate makes the backhaul traffic management very challenging.
 Therefore, network developers should devise novel algorithms and exploit new emerging technologies to overcome traffic management issue \cite{hsu2016resource,wang2014cellular,kabalci20195g,hossain20155g, peng2014system,wong2017key}. Numerous of works in the traffic domain show that the significant traffic is due to downloading popular contents by several users. In addition, re-downloading these contents from the server results in extra backhaul resource consumption leading to the increase of the network cost \cite{li2016cost}. {As a result}, content caching, which caches the popular contents in the storage of BSs, has emerged recently as a promising solution to relieve the burden of the backhaul traffic for improving the QoS of users \cite{li2016cost}. To overcome the storage limitation issues {at} BSs, cooperative caching is shown to be an effective approach \cite{7530876}, where the requested contents, if not available in the serving BS, could be fetched from neighboring BSs (if available in one of them) \cite{ao2017fast}. In this case, the backhaul traffic decreases as more contents are fetched {directly} from the access BSs. Although content caching can decrease the network cost by bringing the contents close to the users, the spectral efficiency and transmit power are the other parameters which are also important from the network cost perspective.
 \\\indent
Another key technology of 5G to overcome the massive connectivity challenge is  non-orthogonal multiple access (NOMA). This technology is envisioned to provide higher spectral efficiency and increase the number of users by sharing the {same} spectrum among {multiple users concurrently} \cite{wei2019performance}. NOMA can be categorized as \textit{1. power domain NOMA} \cite{gurugopinath2020non}, \textit{2. sparse code multiple access (SCMA)} \cite{taherzadeh2014scma}, \textit{3. pattern division multiple access (PDMA)} \cite{chen2019pattern}, \textit{4. low density spreading (LDS)} \cite{mohammed2012performance}, \textit{5. power domain sparse code multiple access (PSMA)} \cite{moltafet2017new} and, \textit{6. lattice partition multiple access (LPMA)} \cite{fang2016lattice}. These techniques are based on the same key concepts, where more than one user is served in each orthogonal resource block, e.g., a time slot, a frequency channel, a spreading code, or an orthogonal spatial degree of freedom \cite{ding2017survey}. In PD-NOMA, different levels of power are assigned to the users who share the same spectrum \cite{gurugopinath2020non}. Applying successive interference cancellation (SIC) at the receiver side is the the main idea of PD-NOMA for signal reception and decoding which is defined extensively in \cite{gurugopinath2020non}.
\vspace{-0.5em}\subsection{Related Works}
Generally, content caching refers to place the popular contents near to the users to enhanced the quality of experience (QoE) \cite{rezaei2016delay}. The authors in \cite{zhang2018hierarchical} investigated different kinds of content caching implementations. Moreover, content placement problem is related to the time and the location of each content which should be cached \cite{zhou2015generalized}. 
 The authors in \cite{peng2014system} provide a basic description about new key technologies for 5G cloud enabled networks and term of cooperative in this work refers to the general form of cooperative resource management such as inter-cell interference coordination and processing pool.
 Also, cooperative content caching is considered widely in \cite{tran2018cooperative, llorca2013network, hoang2018cooperative} to utilize the storage resource.\\
\indent
The authors in \cite{saputra2019distributed, paschos2016wireless, keshavarzian2015clustered, poularakis2014approximation},  investigate the content caching network in an information centric network (ICN) \cite{paschos2016wireless}, a content delivery network (CDN) \cite{paschos2016wireless}, and caching in the edge of wireless networks \cite{saputra2019distributed, keshavarzian2015clustered, poularakis2014approximation}. Although the mentioned papers investigate the content caching network, still the fronthaul links and backhaul links are suffered from heavy loads.
In \cite{tran2018cooperative}, a novel cooperative hierarchical caching framework is proposed to minimize the cost-delay function where delay is considered as a cost factor. Moreover, the authors in \cite{llorca2013network} propose a cooperative caching scheme to minimize the cost, where the cost function factors are related to the traffic load of transformed contents.
\\\indent
The consumed bandwidth cost minimization in cooperative cache management algorithm is considered in \cite{borst2010distributed} while the edge traffic is maximized. The authors in \cite{li2017resource} investigate and propose an efficient resource allocation scheme where a minimum data rate of content delivery is guaranteed in the downlink multiuser non-cooperative cache-enabled orthogonal frequency division multiple access (OFDMA) small-cell networks to maximize the weighted sum of the data rates. The authors in \cite{rezvani2019fairness} design efficient non-cooperative cache placement and delivery strategies for an OFDMA-based cache-enabled HetNet in two separated phases.
  In \cite{7530876}, a cooperative caching in D2D connection scenario is investigated in the caching and delivery phases to minimize the delay of the content delivery process. Besides, \cite{7530876} caches the contents in the caching phase and then delivers the contents in the delivery phase where the radio links are armed by OFDMA.
 \\\indent
  Moreover, two scenarios for NOMA assisted cache-enabled networks are considered in \cite{ding2018noma}. Also, the authors in \cite{8675357} study the cache-enabled NOMA network for two users with the aim of content delivery time minimization. Furthermore, in \cite{8573777}, the downlink resources of D2D networks and edge caching networks are studied, aiming at maximizing the energy efficiency. The caching problem for a heterogeneous small-cell network with bandwidth allocation and caching-aware base station association is proposed in \cite{8399500}. In \cite{8675357}, a cache-aided NOMA scheme for spectral efficiency in downlink transmission is studied. The joint problem of user association and content caching in a dense small-cell is considered in \cite{8598719}.
  \\\indent
 {None of the aforementioned works consider the effect of the network radio resource parameters in the network cost. Although the allocated power to the users in the access links is considered, still the optical links suffer from the huge amount of network traffic due to their limited capacity. In this regard, there is a motivation to study the effect of the consumed resources in the backhaul and the links between two BSs (cooperative links) on the cost function. As it is mentioned before, cooperative content caching can reduce the backhaul traffic while PD-NOMA can deal with the massive connectivity in access links which makes these two technologies available solution for the network traffic  and QoS  challenges. In addition, all of these shortcomings motivate us to propose a novel joint radio management and cooperative caching for PD-NOMA-based HetNets which not only can utilize the network resources and minimize the network cost, but also it can satisfy the required  QoS of the users and network availability for massive machine type communications services \cite{rezvani2019fairness, ding2018noma, 8675357, 8573777, 8598719}.}
  \vspace{-0.5em}\subsection{{Our Contribution and Research Outcomes}}
  In this paper, we consider a network with one macro base station (MBS) and multiple small BSs (SBSs) which can exchange
  contents with each other. We assume that if the requested content does not exist in the storage of the associated BS, the content
  could be delivered by a neighboring  BS or by the backhaul to the serving BS. The main contributions of this paper are
  summarized as follows:
  {
  \begin{itemize}
 \item  In this work, the network cost is optimized not only based on the physical storage of BSs, but also, based on the radio resources in the access links taking into account the consumed resources of the optical links in both fronthaul and backhaul links. We aim to minimize the network cost by considering joint cooperative content caching and PD-NOMA technology. 
  \item In this work, we allocate the backhaul and fronthaul links for each requested content in a joint manner.
  \item We study user association and subcarrier allocation in cooperative cache-enabled networks which results in finding the optimized user assignment to BSs with the optimized usage of resources while providing the requested content in order to minimize the cost. Moreover, we employ PD-NOMA technology in the access links to allocate more than one user to a subcarrier which allows us to reduce the network cost and increases the wireless capacity to support the massive connectivity of future networks. By this way, the considered system model can support the low latency massive connection services stated in Table 1 of \cite{saad2019vision}. {To design near optimal low complexity resource allocation, we adopt the ASM algorithm in which the original problem is divided into three sub-problems (for each phase) each of which is solved iteratively. To solve the resulting problems, we employ the Hungarian algorithm \cite{7530876}, SCA \cite{8976409}, and devise a novel heuristic algorithm with guaranteed convergence. Moreover, via simulation and exploiting the exhaustive search algorithm for a small scale network, we obtain solutions for both original and heuristic algorithms (see Fig. \ref{Optimality_Gap_S}). 
  	\item Numerical results illustrate that the optimality gap of the proposed solution is $6.5$\% on average. Moreover, we proposed a heuristic algorithm to the optimized the content delivery process which the is near optimal. Moreover, our cooperative NOMA scheme outperforms the orthogonal multiple
  	access (OMA) in terms of the delivery cost in general with an acceptable complexity increase. 
  }
   \end{itemize}}
\vspace{-0.5em}\subsection{Paper Organization}
In Section \ref{system model and problem formulation}, we discuss the considered system model. Section \ref{pfsc} and \ref{pfsd} discuss the problem formulation and proposed solution for the caching phase and delivery phase, respectively. Moreover, {in} Section \ref{The Convergence of iterative algorithm and DC approximation}, we prove the convergence and discus about the complexity of employed method.
 In Section \ref{Numerical results}, we verify the performance of proposed network model by numerical results. Also, Section \ref{concolusion} concludes this work.
\vspace{-1em}\section{System Model and Problem Formulation}\label{system model and problem formulation}
\subsection{System Description}
We consider the downlink of a cache-enabled heterogeneous cellular network (HCN) consisting of a single MBS and multiple SBSs. As illustrated in Fig. 1, all BSs are connected to each other via optical fiber fronthaul links. Each BS is in the set of $\mathcal{B}=\{0,1,2,\dots,B\}$ with a size of {$|\mathcal{B}|=B+1$}. The maximum storage capacity for BS $b$ is denoted by $M_b$. {The} MBS is represented by index 0. {We assume that} only one BS provides the service for each user in the set of $\mathcal{U}=\{1,2,\dots,U\}$ with the size of {$|\mathcal{U}|$}. All BSs are connected to a remote content provider (CP) through backhaul links \cite{chen2018explicit}. The unlimited capacity is considered for the backhaul links. The remote CP is assumed to have all $C$ contents denoted by the set $\mathcal{C}=\{1,2,\dots,C\}$.
The size of content $c$, denoted by $s_c$, is modeled by Log-normal distribution with the mean value $\mu_{c}$ and variance $\sigma_c^{2}$. The Zipf distribution is a well-known model for content's popularity \cite{li2016cost, shahid2020energy, lv2020joint, friedlander2019generalization} which is given by
$d_{c} =\frac{1/c^{\alpha}}{\sum_{c=1}^{C} 1/C^{\alpha}} , \forall c \in \mathcal{C}$, where $d_{c}$ and $0 < \alpha \leq 1$ denote the content $c$ popularity and the steepness parameter of Zipf distribution curve, respectively. {The smaller amount of $\alpha$ means the content $c$ is requested with probability by users, and the larger amount of $\alpha$ suppose content $c$ mean low probability of being requested by the users.} 
 We define $\delta_{u,c}(t)\in \{0,1\}$ as the request parameter where $\delta_{u,c}(t)=1$ means user $u$ requests content $c$ at time slot $ t $ and $\delta_{u,c}(t)=0$ otherwise. All the main notations are summarized in Table \ref{notations}.
\begin{table*}[t]
	\centering
	\caption{Table of main notations}
	\label{notations}
	\begin{tabular}{l|l|l|l}
		\hline
		\cellcolor{green}{\textbf{Notations}} & \cellcolor{green}{\textbf{Description}} &\cellcolor{green}{\textbf{Notations}}&\cellcolor{green} {\textbf{Description}} \\
		\hline
		\cellcolor{yellow}{$U$} &\cellcolor{yellow} {Total number of users}&\cellcolor{yellow}	{$\mathcal{B}$} &\cellcolor{yellow}{The set of BSs}\\
		\hline
		\cellcolor{white} {$N$} &\cellcolor{white}  {Number of subcarriers} &\cellcolor{white}{$C$} &\cellcolor{white} {Number of contents}\\
		\hline
		\cellcolor{yellow}{$\alpha$} &\cellcolor{yellow} {Zipf parameter}	&\cellcolor{yellow} {$\sigma^2$} &\cellcolor{yellow} {{Additive} White Gaussian noise power density}\\
		\hline
		\cellcolor{white}{$C_{\text{BW}}$} & \cellcolor{white}{Bandwidth cost constant}&\cellcolor{white}{$C_{\text{Power}}$} & \cellcolor{white}{Power cost constant}\\
		\hline
		\cellcolor{yellow}{$C_{\text{FH}}$} &\cellcolor{yellow} {Cooperative case cost constant}&\cellcolor{yellow}{$C_{\text{BH}}$} &\cellcolor{yellow} {Backhaul case cost constant}\\
		\hline
		\cellcolor{white}{$s_c$} & \cellcolor{white}{Size of content $c$} &\cellcolor{white}{$M_b$} &\cellcolor{white} {Cache storage of BS $b$}\\
		\hline
		\cellcolor{yellow}{$R^{\text{max,\,FH}}_{i,b}$} & \cellcolor{yellow}{Maximum fronthaul rate between BS $i$ and BS $b$}&\cellcolor{yellow}{$r^{\text{FH}}_{i,b,c}$ }&\cellcolor{yellow} {Fronthaul rate variable for content $c$ between BS $i$ and BS $b$}\\
		\hline\cellcolor{white}
		\cellcolor{white}{$r^{\text{BH}}_{b,c}$}
		& 
		\begin{tabular}[c]{@{}l@{}}	
			Backhaul rate variable between remote CP and BS\\ $b$ for content $c$
		\end{tabular}
		&\cellcolor{white}{$y_{i,b,c}(t)$} & \cellcolor{white}{Cooperative delivery case for content $c$ between BS $b$ and $i\,\ i\neq b$}\\
		\hline
		\cellcolor{yellow}{$x^{c}_{b}$} & \cellcolor{yellow}{Cache hit case variable in delivery} &	\cellcolor{yellow}{$z_{b,c}(t)$} & \cellcolor{yellow}{Cache miss case delivery variable}\\
		\hline
		\cellcolor{white}{$r^{n}_{b,u}(t)$} & \cellcolor{white}{Data rate for user $u$ in BS $b$ on $n^{\text{th}}$ subcarrier} &\cellcolor{white}	{$p^{n}_{b,u}$} & \cellcolor{white}{Transmission power for user $u$ in BS $b$ on $n^{\text{th}}$ subcarrier}\\
		\hline
		\cellcolor{yellow}{$\tau^{n}_{b,u}(t)$ }& \cellcolor{yellow}
		\begin{tabular}[c]{@{}l@{}}	
			Subcarrier assignment for user $u$ in BS $b$ on\\ $n^{\text{th}}$ subcarrier 
		\end{tabular}
		&\cellcolor{yellow}	{$\gamma^{n}_{b,u}$} & \cellcolor{yellow}{SINR parameter for user $u$ in BS $b$ on $n^{\text{th}}$ subcarrier}\\
		\hline
		\cellcolor{white}{$h^{n}_{b,u}$ }&\cellcolor{white}{ Channel gain }&\cellcolor{white}{$W^{\text{tot}}$} & \cellcolor{white}{Total radio bandwidth}\\
		\hline
		\cellcolor{yellow}{$W$}&\cellcolor{yellow} {Subcarrier bandwidth} &\cellcolor{yellow}{$d_c$} & \cellcolor{yellow}{Content popularity for content $c$}\\
		\hline
		\cellcolor{white}$\delta_{u,c}(t)$&\cellcolor{white}{ Requesting parameter for content $c$ by user $u$} &\cellcolor{white}{$\rho_{b,c}(t')$} &\cellcolor{white}{Content placement variable for content $c$ on BS $b$}\\
		\hline
		\cellcolor{yellow}{$I^{n,\text{Intra}}_{b,u}$}&\cellcolor{yellow}{ Intra interference} & \cellcolor{yellow}{$I^{n,\text{Inter}}_{b,u}$}& \cellcolor{yellow}{Inter interference}\\
		\hline
		\cellcolor{white}{$P_{b}^{\text{max}}$} & \cellcolor{white}{Maximum power of BS $b$} &\cellcolor{white}{$P_b^{\text{mask}}$} &\cellcolor{white}{ Maximum power on subcarriers in BS $b$}\\
		\hline
		\cellcolor{yellow}{$\bar{\Phi}_{b}$} & \cellcolor{yellow}{Cooperative case and Cache miss case cost}&	\cellcolor{yellow}{$l_{b}^{\text{max}}$} & \cellcolor{yellow}{Maximum number of user assignment in one subcarrier in BS $b$}\\
		\hline
		\cellcolor{white}{$T$} & \cellcolor{white}{Content delivery process time period} &  \cellcolor{white}{$\Phi^{\text{tot}}$} &\cellcolor{white} {Total cost of network}\\
		\hline
	\end{tabular}
\end{table*}
 As shown in Fig. \ref{framework}, we presume two time phases in our network; the first phase refers to cache-placement and the second is about content delivery. Hereafter, the delivery phase is divided into several small time slots {where} all the users' requests are supplied by the BSs. We suppose that each user  requests for only one content in each small time slot \cite{rezvani2019fairness}.

\vspace{-0.5em}\begin{figure}
	\centering
	\includegraphics[width=0.5\textwidth]{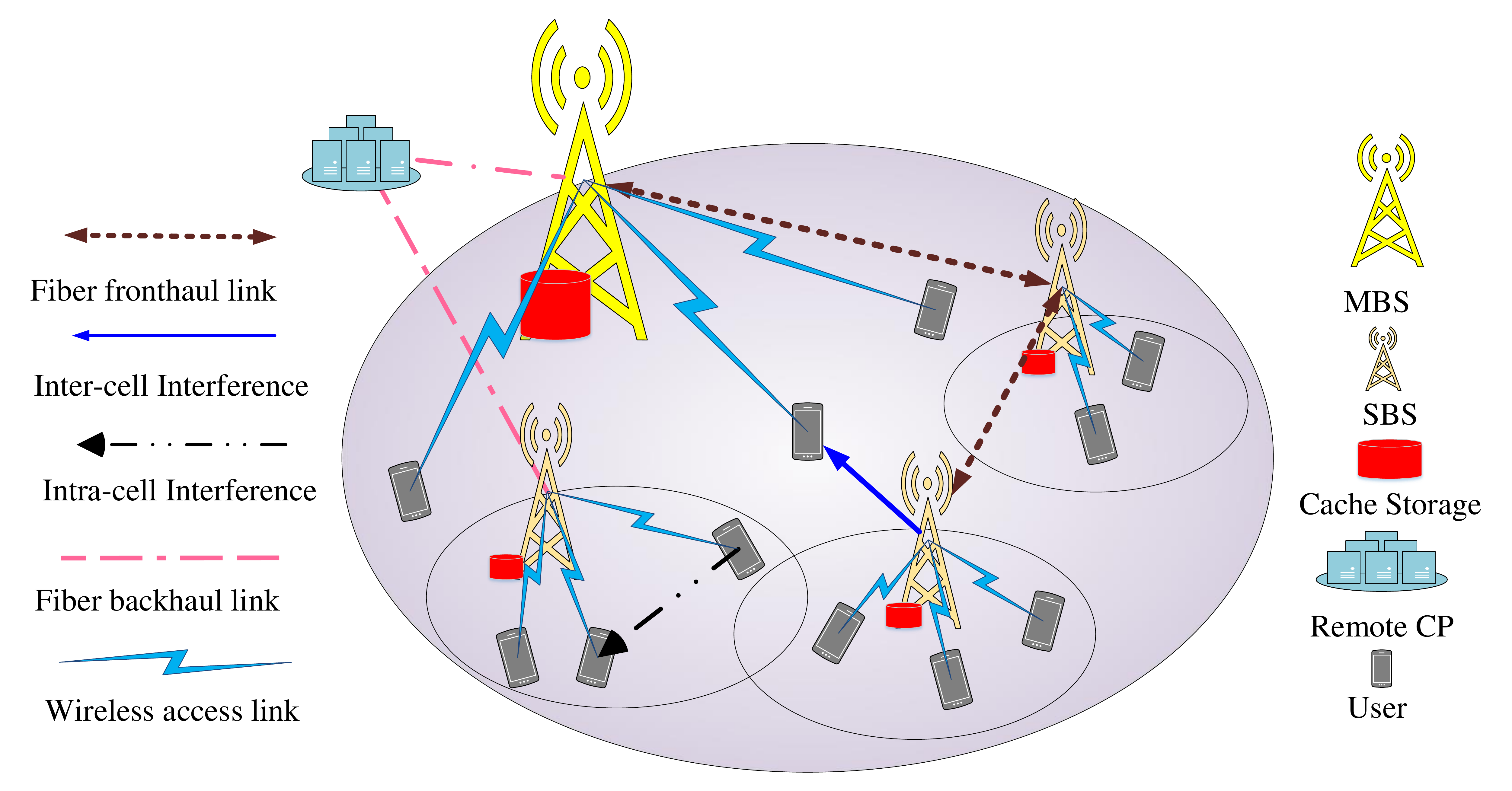}
	\vspace{-0.5em}\caption{The proposed network model with one MBS and multiple SBSs.} 
\end{figure}

\begin{figure}\centering
\includegraphics[scale=0.17]{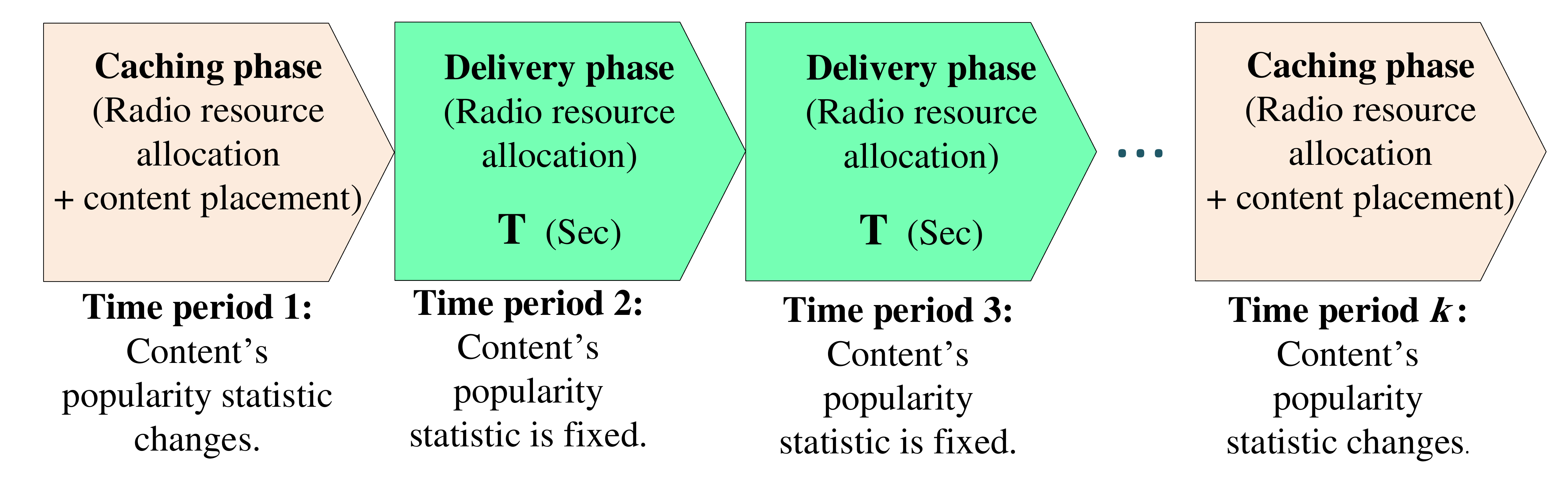}
\centering \caption{Definition for caching phase and delivery phase.}
\label{framework}
\end{figure}

\vspace{-0.75em}\subsection{\textbf{Caching Phase}}
{We consider a continuous long period of time in the set $t'\in\mathcal{V'}=\{1, 2, \dots,V'\}$ when the content popularity statistic is fixed \footnote{In our proposed network, the caching phase executes during off-peak time and is valid for long duration as the popularity distribution information (PDI) changes slowly and remains constant for a long period of time \cite{8611393, li2015distributed, cheng2017power}. In this regard, we assume that the next caching phase occurs when the PDI is changed. During this long period of time (till next caching phase), operators can monitor and collect the number of users' requests for each content and obtain the PDI for the next caching phase \cite{8611393}. In this regard, the PDI is constant for the next delivery phase \cite{ao2017fast, 8611393}.}. Until next caching phase the time period is slotted into several equal mini delivery time slots with specific duration $T$ which is shown in Fig. 2. The mini time slots are assumed to be in the set $t\in\{1, 2,\dots, V\}$.} The caching phase occurs in low traffic hours of the network where the contents are cached through the backhaul \cite{rezvani2019fairness}. The {popularity distribution information (PDI)} is fixed for a long time and the caching phase repeats when the PDI changes. Moreover, in our considered network, the content placement is accomplished with ergodic {(which means the average value of all cases)} radio allocation of resources with the PDI and channel distribution information (CDI) as the inputs \cite{rezvani2019fairness}. After the caching phase, all BSs are ready to deliver contents to their users.
\vspace{-0.75em}\subsection{{\textbf{Delivery Phase}}}
\begin{figure}[t]
\centering
\includegraphics[scale=0.21]{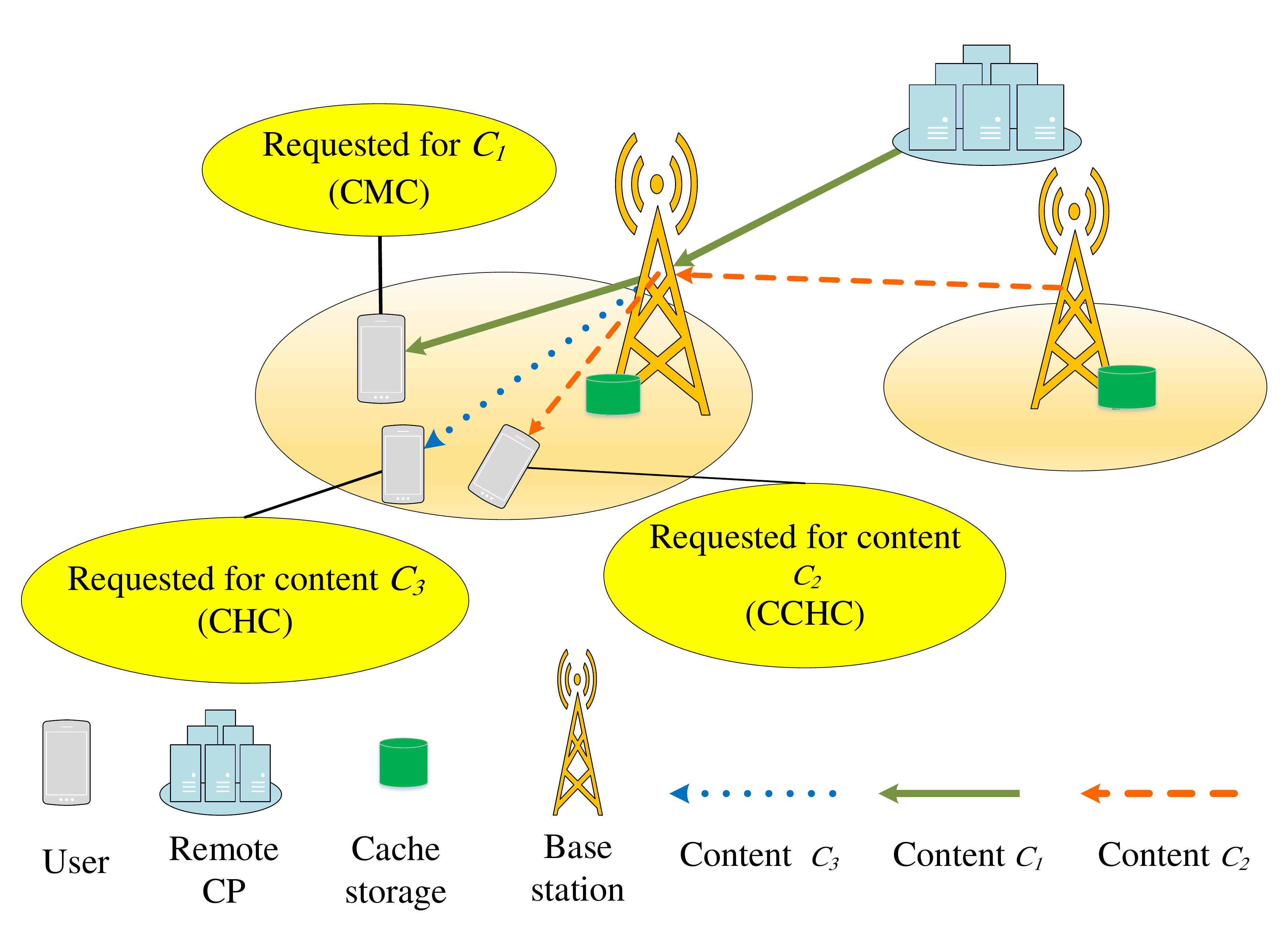}
\centering \caption{Our proposed content delivery scheme for three possible cases.}
\label{parallel}
\end{figure}

After {accomplishing} the caching phase, the delivery phase is launched. We assume that over every short time slot, users location and the channel gains of the access network are fixed. In addition, it is assumed that the channel gains in each short time slot are independent from each other. After receiving content requests, the BSs begin to transmit the requested contents to their users.
 The procedure of providing the requested contents is defined in the following:
\begin{enumerate}
\vspace{-0.5em}\item
As BS $b$ receives all the requests from its users, it checks its storage to find whether the requested content $c$ exists in its storage or not. If the content $c$ exists in the storage of $b$, the cache hit case (CHC) occurs as Fig. \ref{parallel}. In particular, $c$ is sent immediately to the users who requested for it.
\item
Otherwise, BS $b$ sends a request for content $c$ to all of the BSs in neighbor. Each BS $i$ ($i\neq b$), which receives the request from BS $b$, searches for content $c$ in its storage. From all the BSs which have content $c$ in their storage, BS $i$ is selected to send content $c$ to BS $b$ via fronthaul fiber link. Upon getting content $c$, BS $b$ sends it to all the users who request content $c$. In this case, cooperative cache hit case (CCHC) occurs which is shown in Fig. \ref{parallel}.
\item
If the requested content $c$ is not cached in any BSs, BS $b$ forwards the request to the remote CP for fetching content $c$. Then, the content is transmitted by the remote CP to BS $b$ and the procedure terminates after content $c$ is disseminated to the users by the associated BS $b$. In this case, cache miss case (CMC) happens as illustrated in Fig. \ref{parallel}. Note that content $c$ is delivered from the remote CP once to each BS $b$.
\end{enumerate}
	We denote by $W^{\text{tot}}$ the total downlink bandwidth which is divided into $N$ orthogonal subcarriers with {a} subcarrier bandwidth of $W=\frac{W^{\text{tot}}}{N}$. Additionally, the achievable data rate at the time slot $ t $ of the access link for user $u$ from BS $b$ over the $n^{\text{th}}$ subcarrier can be calculated by
$r_{b,u}^{n}(t)= W \cdot \log_{2}[1+\tau_{b,u}^{n}(t)\cdot \gamma_{b,u}^{n}(t)], \,\ \forall b\in\mathcal{B}, n\in\mathcal{N},
 u\in\mathcal{U},$
where $\tau_{b,u}^{n}(t)\in\{0,1\}$ is the subcarrier assignment parameter with $\tau_{b,u}^{n}(t)=1$ if the $n^{\text{th}}$ subcarrier is assigned to user $u$ in BS $b$ at time slot $ t $, and $\tau_{b,u}^{n}(t)=0$ otherwise. Also, $\gamma_{b,u}^{n}(t)$ is used to denote the received signal-to-interference-plus-noise ratio (SINR) of user $u$ from BS $b$ over subcarrier $n$ at time slot $ t $ which is given by
$\gamma_{b,u}^{n}(t)=\frac{p_{b,u}^{n}(t)\cdot h_{b,u}^{n}(t)}{I_{b,u}^{n,\text{Intra}}(t)+I_{b,u}^{n,\text{Inter}}(t)+\sigma^2}, \,\ \forall u\in\mathcal{U}, b\in\mathcal{B},
 n\in\mathcal{N}$, 
where $p_{b,u}^{n}(t)$ is the transmit power from BS $b$ to user $u$ on the $n^{\text{th}}$ subcarrier. In addition, $h_{b,u}^{n}(t)$ is considered to be the channel power gain between user $u$ and BS $b$ over subcarrier $n$. Also, $\sigma^2$ is the power of the additive white Gaussian noise (AWGN). Furthermore, 
$I_{b,u}^{n,\text{Intra}}(t)=\sum\limits_{\scriptstyle i \ne u\atop{\scriptstyle i\in\mathcal{U} \atop\scriptstyle h_{b,i}^n(t)\geq h_{b,u}^n(t)}}\tau_{b,i}^{n}(t)\cdot p_{b,i}^{n}(t)\cdot h_{b,u}^{n}(t),$
is the received interference of user $ u $, which is assigned to bs $ b $, on subcarrier $ n $ from all other users of bs $ b $ which are assigned the same subcarrier based on NOMA. In this case, if user $i$ has a larger channel gain than user $u$, it produces as an interference on user $u$. Also, 
$I_{b,u}^{n,\text{Inter}}(t)=\sum\limits_{\scriptstyle j\in\mathcal{B}\atop{\scriptstyle j\ne b}}\sum_{d\in\mathcal{U}}\tau_{j,d}^{n}(t)\cdot p_{j,d}^{n}(t)\cdot h_{j,u}^{n}(t),$
 is the interference of users from BS $j$ on user $u$. 
 Note that the content delivery process of content $c$ in each BS $b$ should be done just by one of the delivery
cases (CHC, CCHC, or CMC). As shown in Fig. \ref{parallel}, there is no buffer to save content $c$ from CCHC and CMC.
Hence, content $c$ is disseminated to the users who request for it as it is fetched from BS $i$ or the remote CP.
%
\vspace{-0.75em}\section{\textbf{Problem Formulation and Solution of the Caching Phase}}\label{pfsc}
\vspace{-0.5em}\subsection{Objective Function of The Caching Phase}
{ As the cost of the network is one of the most important factors for the 
 	communication service providers, our aim is cost minimization.
The considered cost comprises of three terms as: the cost of the utilized power, radio bandwidth, and transmission bandwidth as follows:
{\paragraph*{\textbf{\textit{Power Consumption}}} In general, the  total consumption power at each BS $ b $ at time slot $ t $ comprises of the communication and computing power at baseband  unit (BBU) and can be modeled as follows \cite{7510703, 7926940}:
	\begin{align}\label{Power_T1}
	P_{b}^{\text{Total}}(t)=P_{b}^{\text{Comm}}(t)+P_{b}^{\text{BBU}},
	\end{align}
	{where}
	\begin{align}
	&P_{b}^{\text{Comm}}(t)=
	\begin{cases}
	{P_{b}^{\text{Hardware}}}+P_{b}^{\text{TX}}(t), & {0<P_{b}^{\text{TX}}(t) \leq P_b^{\max}}, \\
	{P_{b}^{\text{Sleep}}}, & {P_{b}^{\text{TX}}(t)=0},
	\end{cases}
	\end{align}
	where $ P_{b}^{\text{TX}}(t)=\sum_{u=1}^{U}\sum_{n=1}^{N}C_{b}^{\text{Power}} p_{b,u}^{n}(t)$ and $C_{b}^{\text{Power}}(\frac{\text{price}}{\text{mWatts}})$ is the unit cost for the transmit power. Hence, the first part of cost is 
	$ \phi^{\text{Power}}(t)= \sum_{b=0}^{B}P_{b}^{\text{Total}}(t).$
	{More details about the power consumption  in the radio part and modeling for  multiple antennas}  transmission can be found in \cite{7926940} and \cite{8330764}, respectively.
%
}
\paragraph*{\textbf{\textit{Radio Bandwidth Consumption}}}
The second part of the cost is the radio bandwidth consumption cost at each BS at time slot $ t $ obtained as
\begin{align}
\phi^{\text{BW-Radio}}(t)=\sum_{b=0}^{B}\sum_{u=1}^{U}\sum_{n=1}^{N}C^{\text{BW}}\cdot \tau^{n}_{b,u}(t)\cdot W,
\end{align}
where constant  $C^{\text{BW-Radio}}(\frac{\text{price}}{\text{KHz}})$  is the unit cost of consumption of radio bandwidth.}
\paragraph*{\textbf{\textit{Storage Constraint}}}
The content placement in the BS's storage depends on the storage capacity. In other words, the total size of the cached contents in the storage of BS should not  exceed the storage capacity. Therefore, we impose the following constraint:
\begin{align}\label{cache_size_constraint}
\sum_{c=1}^C \rho_{b,c}(t')\cdot s_c \leq M_b , \forall b\in\mathcal{B},~t'\in\mathcal{V'},
\end{align}
{where $\rho_{b,c}(t')\in \{0,1\}$ is the content placement variable, where $\rho_{b,c}(t')=1$ means  content $c$ exists in BS $b$, otherwise $\rho_{b,c}(t')=0$. As our cached contents are not changed till time slot $t'+1$, for every mini delivery time slot $t$, the content placement variable has the same value. In this case for simplicity, we write $\rho_{b,c}(t')$ as $\rho_{b,c}$.}
{
\paragraph*{\textbf{\textit{Power Constraints}}}
The allocated transmit power for all users in BS $b$ should not exceed the maximum power budget as stated in the following constraint:
\begin{align}\label{dp5}
\sum_{u=1}^{U}\sum_{n=1}^{N} \tau_{b,u}^{n}(t)\cdot p_{b,u}^{n}(t) \leq P_{b}^{\text{max}}, \,\ \forall  b\in\mathcal{B},
\end{align}
where $P_{b}^{\text{max}}$ is the maximum power budget in BS $b$. Also, there is a maximum power limitation for each subcarrier as follows:
\begin{align}\label{dp6}
0\leq p_{b,u}^{n}(t)\leq P_{b}^{\text{mask}}, \,\ \forall u\in\mathcal{U}, b\in\mathcal{B}, n\in\mathcal{N},
\end{align}
where $P_{b}^{\text{mask}}$ is the maximum transmit power BS $b$ on each subcarrier $n$.
\paragraph*{\textbf{\textit{PD-NOMA Constraints}}}
The number of the assigned users to one subcarrier is restricted as follows:
\begin{align}\label{dp8}
\sum_{u=1}^{U}\tau_{b,u}^{n}(t) \leq l_{b}^{\text{max}}, \,\ \forall n\in\mathcal{N}, b\in\mathcal{B},
\end{align}
where $l_{b}^{\text{max}}$ is the maximum allowed number of assigned users to one subcarrier. Moreover, the SIC in PD-NOMA is assured by
\begin{align}\label{dp7}
&
\frac{h_{b,u}^{n}(t)}{I_{b,u}^{n,\text{Intra}}(t)+I_{b,u}^{n,\text{Inter}}(t)+\sigma^2} \geq
\frac{h_{b,u'}^{n}(t)}{I_{b,u'}^{n,\text{Intra}}(t)+I_{b,u'}^{n,\text{Inter}}(t)+\sigma^2}, \,\ \nonumber \\
&
\forall b\in\mathcal{B}, u, u' \in \mathcal{U},u'\neq u ,h_{b,u}^{n}(t) \geq h_{b,u'}^{n}(t),
n\in\mathcal{N}.
\end{align}
{This constraint {guarantees} that user $u$ can detect and decode the signal of user $u'$ and can remove it from the received signal.}
\paragraph*{\textbf{\textit{BS Assignment Constraint}}}
In this paper, it is presumed that each user is assigned to one BS. Therefore, we have
{\begin{align}\label{dp9}
\tau_{b,u}^{n}(t)+\sum\limits_{\scriptstyle b' = 1\hfill\atop
\scriptstyle b' \ne b\hfill}^{B}\tau_{b',u}^{n'}(t)\leq 1,\forall b\in\mathcal{B}, n, n'\in\mathcal{N}, u\in\mathcal{U}.
\end{align}
}
\vspace{-0.75em}\subsection{\textbf{Caching Phase}}\label{g cachshode}
The following optimization problem should be solved for determination of the content placement:
\begin{subequations}
\label{cp}
\begin{align}
&
\minimize_{\boldsymbol{p, \tau, \rho}}\Bbb{E}_{\bold{H}(t)}\{\phi^{\text{Power}}(t)+\phi^{\text{BW-Radio}}(t)\} \nonumber\\
&
\text{s.t. }\nonumber\\
&
\label{cp5}
\sum_{u=1}^{U}\sum_{n=1}^{N} \Bbb{E}_{\bold{H}(t)}\{\tau_{b,u}^{n}(t)p_{b,u}^{n}(t)\} \leq P_{b}^{\text{max}}, \,\ \forall  b\in\mathcal{B},\\
&
\label{cp6}
0\leq p_{b,u}^{n}(t)\leq P_{b}^{\text{mask}}, \,\ \forall u\in\mathcal{U}, b\in\mathcal{B}, n\in\mathcal{N},
\\
&
\nonumber
\label{cp7}
\Bbb{E}_{\bold{H}(t)}\{\frac{h_{b,u}^{n}(t)}{I_{b,u}^{n,\text{Intra}}(t)+I_{b,u}^{n,\text{Inter}}(t)+\sigma^2}\} \geq
\\
&
\Bbb{E}_{\bold{H}(t)}\{\frac{h_{b,u'}^{n}(t)}{I_{b,u'}^{n,\text{Intra}}(t)+I_{b,u'}^{n,\text{Inter}}+\sigma^2}(t)\}, \,\ \nonumber\\
&
\forall b\in\mathcal{B}, u, u' \in \mathcal{U},u'\neq u ,h_{b,u}^{n}(t) \geq h_{b,u'}^{n}(t), n\in\mathcal{N},\\
&
\label{cp8}
\sum_{u=1}^{U}\tau_{b,u}^{n}(t) \leq l_{b}^{\text{max}}, \,\ \forall n\in\mathcal{N},  b\in\mathcal{B},\\
&
\label{cp9}
\tau_{b,u}^{n}(t)+\sum\limits_{\scriptstyle b' = 1\hfill\atop
	\scriptstyle b' \ne b\hfill}^{B}\tau_{b',u}^{n'}(t)\leq 1,\forall b\in\mathcal{B}, n, n'\in\mathcal{N}, u\in\mathcal{U},
\\
&
\label{cp10}
\sum_{c=1}^{C}d_{c}\cdot\rho_{b.c}\cdot s_{c} \leq\sum_{u=1}^{U}\sum_{n=1}^{N}T\Bbb{E}_{\bold{H}(t)}\{r^{n}_{b,u}(t)\}, \,\ \forall b\in\mathcal{B}, \\
&
\label{cp11}
\sum_{c=1}^C \rho_{b}^{c}s_c\leq M_b, \,\ \forall b\in\mathcal{B},\\
&
\label{cp15}
\rho_{b,c}\in \{0,1\}, \,\ \forall b\in\mathcal{B}, c\in\mathcal{C},\\
&
\label{cp16}
\tau_{b,u}^{n}(t)\in \{0,1\}, \,\ \forall b\in\mathcal{B}, u\in\mathcal{U}, n\in\mathcal{N},
\end{align}
\end{subequations}
where
 $\bold{p}=[p_{b,u}^{n}(t)],\,\boldsymbol{\tau}=[\tau_{b,u}^{n}(t)],\,\boldsymbol{\rho}=[\rho_{b,c}],~\bold{H}(t)=[h_{u,b}^{n}(t)]$, and $\Bbb{E}_{\bold{H}(t)}\{\cdot\}$ is the expectation over $\bold{H}(t)$. {Note that caching is usually performed in a large time scale, e.g. one or few times per day. In order to consider all fading states, we consider average-based content caching instead of instantaneous caching in \eqref{cp} \cite{rezvani2019fairness}.}
 Constraints 
\eqref{cp5} and \eqref{cp6} are the maximum power budget in each BS $b$ and on each subcarrier $n$ in each BS $b$, respectively.
Also, \eqref{cp7} ensures the successful SIC in PD-NOMA. Moreover, \eqref{cp8} limits the maximum number of users allocated to each subcarrier. \eqref{cp9} shows that each user can be served just by one BS. \eqref{cp10} ensures that all the cached contents in BS $b$ with popularity $d_c$ can be delivered in $T$ seconds with the ergodic rate for all users in BS $b$.
 Obviously, the caching phase problem is non-convex and mixed-integer non-linear problem (MINLP). Therefore, existing convex optimization methods {are not applicable to the problem at hand, as an alternative, we aim to design a new computationally-efficient sub-optimal algorithm based on ASM}. Based on this method, the caching phase problem is divided into three sub-problems, namely content placement problem in which we find $\boldsymbol{\rho}$, subcarrier assignment problem in which we find $\boldsymbol{\tau}$, and transmit power and data rate sub-problem in which the resource parameters are found. The proposed ASM based algorithm for the caching phase and the delivery phase is stated in Alg. 1. {Also, the overview of proposed algorithms for the caching phase is shown in Fig. 4. The initial point of the caching phase can be found on Appendix A.}
\begin{algorithm}
\caption{{The proposed ASM for the caching phase}}
\begin{algorithmic}[1]\label{ASMCACH}
\STATE For $s=1$ to $S$ which is the maximum number of iteration, and $0<\Upsilon<<1$ is the tolerance (or accuracy):
\STATE \textbf{repeat}
\STATE~~Find $\boldsymbol{\rho}^{s}$ by solving \eqref{rhocp} via utilizing MOSEK Toolbox.
\STATE~~Find $\boldsymbol{\tau}(s)$ by solving \eqref{taucp} via using the Hungarian algorithm  which is stated in Alg. 2.
\STATE~~Find $\bold{p}^{s}$ by solving \eqref{ro v tau sabet} via applying SCA-DC Approximation which is presented in Alg. 3.
\STATE  $S=s+1$
\STATE \textbf{Until} $|\phi^{\text{tot}}(s)-\phi^{\text{tot}}(s-1)|$$\le$$\epsilon$ or $s=S$.
\STATE \textbf{Output} $\boldsymbol{\tau},\,\boldsymbol{\rho},\,\bold{p},\,\bold{r}^{\text{FH}},\,\bold{r}^{\text{BH}}$.
\end{algorithmic}
\end{algorithm}
\begin{figure}\label{fchcache}
	\centering
	\includegraphics[width=0.37\textwidth]{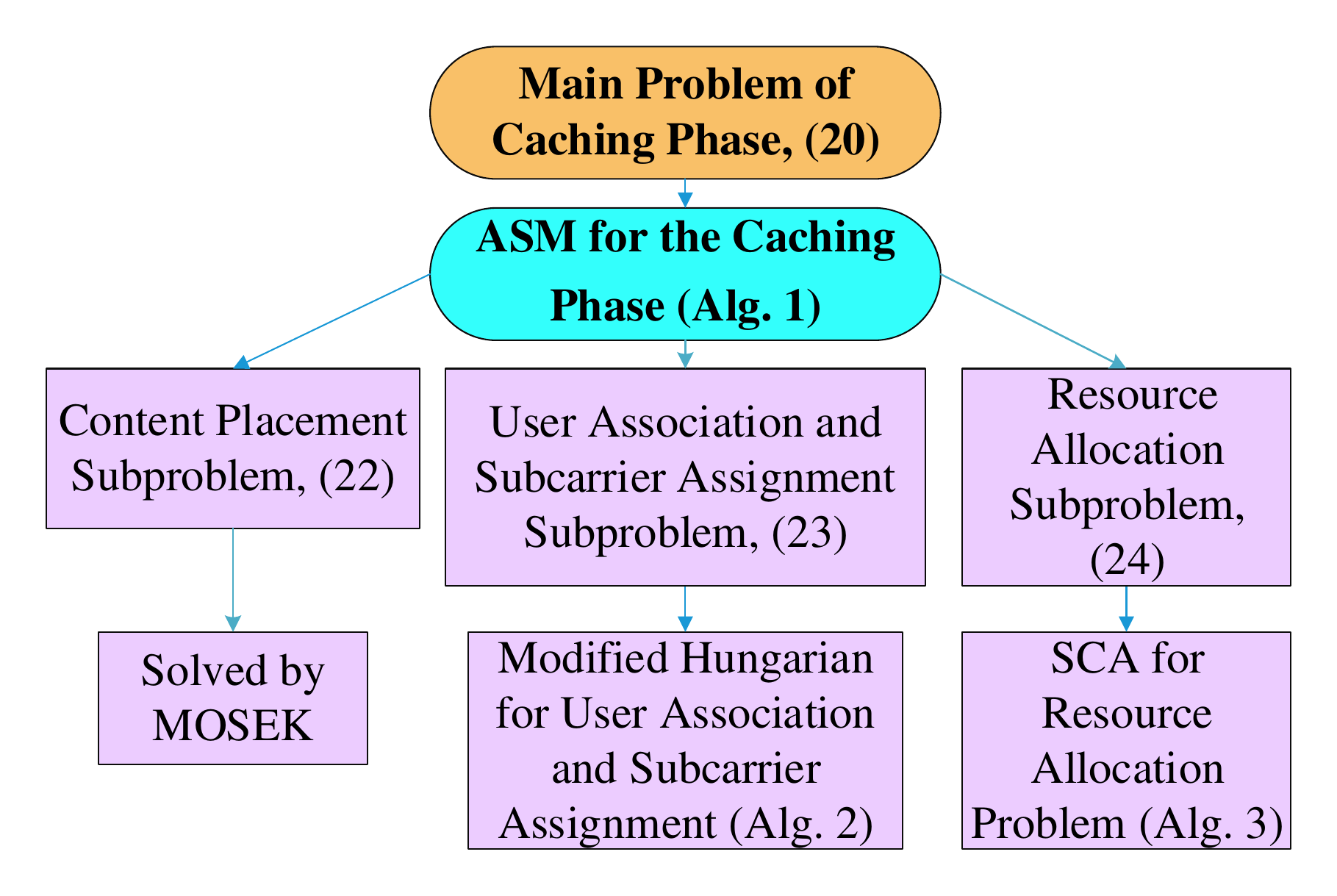}
	\vspace{-0.75em}\caption{The overview of the proposed algorithm in the caching phase.}
\end{figure}\vspace{-0.5em}
\vspace{-0.5em}\subsubsection{\textbf{Content Placement Sub-problem and Solution}}\label{contentplacement}
The first sub-problem is {to find}  $\boldsymbol{\rho}$, which is:
\begin{align}\label{rhocp}
&
\minimize_{\boldsymbol{\rho}}\Bbb{E}_{\bold{H}(t)}\{\phi^{\text{Power}}(t)+\phi^{\text{BW-Radio}}(t)\}\\
&
\text{s.t. }~\nonumber
\eqref{cp10}, \eqref{cp11}, \eqref{cp15}. \nonumber
\end{align}
 The mentioned problem is {a standard} integer linear programming (ILP) problem which is solved by MOSEK solver toolbox in MATLAB.
\vspace{-0.5em}\subsubsection{\textbf{User Association and Subcarrier Assignment Sub-problem in Caching Phase}} \label{2nd sp & s}
To find $\boldsymbol{\tau}$, the following non-convex problem is solved:
\begin{align} \label{taucp}
&
\minimize_{\boldsymbol{\tau}}\Bbb{E}_{\bold{H}(t)}\{\phi^{\text{Power}}(t)+\phi^{\text{BW-Radio}}(t)\}\\
&
\text{s.t. }~\nonumber
 \eqref{cp5}, \eqref{cp7}, \eqref{cp8}, \eqref{cp9}, \eqref{cp10},
\eqref{cp16}. \nonumber
\end{align}
The Hungarian algorithm \cite{7530876, 7762733} is applied for \eqref{taucp}. Furthermore, the defined subcarrier parameter shows the user association and subcarrier allocation, jointly. In this case, the Hungarian algorithm is applied two times in this problem (for user association and subcarrier allocation). The {sub-optimal solution} for \eqref{taucp} is depended on the minimum transmission power allocation, aim to the network cost minimization. Regard to this, $\bold{P}$ is considered to be the cost matrix of our Hungarian algorithm which should be minimized. {In this case, the sub-problem is divided into the user association part and the subcarrier allocation part. At the beginning, we define the Hungarian algorithm for the user association problem and then we describe the Hungarian algorithm for the subcarrier allocation.}
\vspace{-0.5em}\subsubsection*{{\underline{\textit{User Association}}}}
As matrix $\bold{P}_1=[p_{b,u}]$ is not a square matrix, the problem is an unbalanced problem for user association. To make $\bold{P}_1$ as a square matrix, we consider $\mathcal{B}'$ as a set of virtual BSs which $|\mathcal{B}'|= ||\mathcal{U}|-|\mathcal{B}||$ (we assume that the users' number is more than the number of BSs). Moreover, $\mathcal{\Psi}=\mathcal{B}'\cup\mathcal{B}$ is defined. In this regard, a square matrix $\bold{P}'_1=\{p_{u,m}|u=1,2,\dots,|\mathcal{U}|\ , m=1,2,\dots,|\mathcal{\Psi}|\}$ is defined to change the unbalanced user association problem to the balanced one. Also, each element of $\mathcal{B}'$ which is connected to any elements of  $\mathcal{U}$ is equaled to 0.
\begin{description}
          \item[\textbf{Step 1}:]~~Reduction of each row $u=1,2,\dots,|\mathcal{U}|$, $P'_{{1}_{u,m}}-\min \{P'_{{1}_{u,m}}|u=1,2,\dots,|\mathcal{U}|\}$.
          \item[\textbf{Step 2}:]~~Reduction of each column  $m=1,2,\dots,|\mathcal{\Psi}|$, $P'_{{1}_{u,m}}-\min \{P'_{{1}_{u,m}}|m=1,2,\dots,|\mathcal{\Psi}|\}$.
          \item[\textbf{Step 3}:]~~In this step, all the zeros should be covered by the minimum number of lines.
          \item[\textbf{Step 4}:]~~Check for accomplishment of optimal assignment. If the minimum number of essential lines for covering zeros equals to the number of rows, an optimal set of assignment is feasible. In this case, go to \textbf{Step6}. Otherwise go to the next step.
          \item[\textbf{Step 5}:]~~If the number of essential lines for covering all zeros is less than the number of the rows, modify $\boldsymbol{P}$ in the following way:
\begin{description}
  \item[a.]Subtracting the smallest uncovered element from every uncovered element in $\boldsymbol{P}$.
  \item[b.]The smallest uncovered element should be added to the covered elements of each column.
  \item[c.]The elements which did not change, should be the same as the first.
\end{description}
Repeat Steps 4 and 5 until an optimal set of assignment is achieved.
\item[\textbf{Step 6:}]~~Apply the assignment in the positions that have zero elements once at a time.
Begin with the row or column that has only one zero. Each row and each column need to receive exactly one assignment, hence, both of the row and the column, which is involved, should be crossed out after each assignment is made. By preferring the rows and columns which has fewer zeros, the algorithm continues with the rows and columns which are not cross out yet to select the next assignment. The process needs to go on until every row and every column have been crossed out, i.e., they have exactly one assignment.
        \end{description}
\subsubsection*{\textit{\underline{Subcarrier Allocation}}}
To determine the allocated subcarrier for each user $u$, the Hungarian algorithm is applied for allocating the subcarriers of each BS $b$ to their users. In this case, we define $\mathcal{U}_b=\{1,\dots,u_b\}$ as the set of users in BS $b$. Also, $\bold{P}_2=[p_{u_b}^{n}]$ is not the square matrix too. To deal with non-squareness of matrix $\bold{P}_2$, sets of $|\mathcal{N}'|= ||\mathcal{U}_b|- |\mathcal{N}||$ and $\mathcal{\chi}=\mathcal{N}'\cup\mathcal{N}$ are defined (we assume that the number of users are more than the number of subcarriers in each BS $b$). Furthermore,
a square matrix $\bold{P}'_2=\{p_{u,q}|u=1,2,\dots,|\mathcal{U}_b|\ , q=1,2,\dots,|\mathcal{\chi}|\}$ is introduced.
Then, all the steps of the Hungarian algorithm is applied to $\bold{P}'_2$ to assign all users in BS $b$ to the
subcarriers.\\
After applying the Hungarian algorithm for subcarrier allocation problem, all the constraints should be checked. Note that, \eqref{cp10} may not {be} satisfied for all of the users in BS $b$. To deal with this problem, for all the users in
$\mathcal{U}''$ (the set of users who the constraint \eqref{cp10} is not satisfied for them), we recast matrix
$\boldsymbol{P}''=\{P_{u',n}=|u'=1,2,\dots,|\mathcal{U}''|\ , n=1,2,\dots,|\mathcal{N}|\}$ for $l_b^{\text{max}}$ times.
Then, we choose those minimum elements of the mentioned matrix, which can satisfy \eqref{cp10} via first answers
of subcarrier assignment. Hence, by applying this method, we solve \eqref{taucp}. {The overall algorithm is summarized in Alg. 2.}
\begin{algorithm*}[tp]\label{hungarian}
\caption{{Modified Hungarian-based algorithm for user association and subcarrier assignment in
caching phase}}
\begin{algorithmic}
\STATE\textbf{\underline{User Association:}}
\STATE \textbf{Input}: $\bold{P}_1=[p_{b,u}]$
\STATE \textbf{Output}: $\boldsymbol{\tau}_1=[\tau_{b,u}]$
\STATE Define a set $\mathcal{B}'$ with number of virtual BSs $||\mathcal{U}|-|\mathcal{B}||$.
\STATE Define a set $\mathcal{\Psi}=\mathcal{B}'\cup\mathcal{B}$.
\STATE Define a square matrix $\bold{P}'_1=\{p_{u,m}|u=1,2,\dots,|\mathcal{U}|\ ,m=1,2,\dots,|\mathcal{\Psi}|\}$
\STATE Calculate the cost matrix $\bold{P}'_1$ :
\begin{itemize}
\item i) The cost value of link $u\in \mathcal{U}$ to $b\in \mathcal{B}$ is $p_{b,u}$;
\item ii) The cost value of each element of set $\mathcal{B}'$ connected to any element of $\mathcal{U}$ is 0.
\end{itemize}
\STATE Use the proposed Hungarian algorithm to solve the assignment problem with cost matrix $\bold{P}'_1$ and obtain the optimal assignment solution.
\begin{itemize}
\item i)  Row reduction for $\bold{P}'_1$.
\item ii)  Column reduction for $\bold{P}'_1$.
\item iii)  Test whether an optimal assignment can be completed by determining the minimum number of lines needed to cover all zeros in ${\bold{P}'_1}$. If the number of lines equals the number of rows, an optimal set of assignment is feasible. In case, go to v). Otherwise go to next step.
\item iv)  If the number of lines is less than the number of rows, modify ${\bold{P}'_1}$ as follows: \\
a. Subtract the smallest uncovered element from every uncovered elements in ${\bold{P}'_1}$.\\
b. Add the smallest uncovered element to those elements at intersections of covering lines.\\
c. The elements which are covered but not at the intersections of covering lines remain unchanged in the next update.\\
Repeat iii) and iv) until an optimal set of assignment is feasible.
\item v)  Make the assignments once at a time in positions that have zero elements.
\end{itemize}
\STATE Map the obtained solution to the assignment decision matrix $\boldsymbol{v}'$ which equals to $\boldsymbol{v}'=[\tau_{b,u}]$.
\STATE
\STATE \textbf{\underline{Subcarrier Assignment:}}
\STATE Then for each BS $b, \forall b\in\mathcal{B}$:
\STATE \textbf{Input}:$\bold{P}_2=[p_{u_b}^{n}]$
\STATE \textbf{Output}:$\tau_{u_b}^{n}$
\STATE Define a set $\mathcal{N}'$ with number of virtual subcarriers $||\mathcal{U}_b|-|\mathcal{N}||$.
\STATE Define a set $\mathcal{\chi}=\mathcal{N}'\cup\mathcal{N}$.
\STATE Define a square matrix $\bold{P}'_2=\{p_{u,q}|u=1,2,\dots,|\mathcal{U}_b|\ ,q=1,2,\dots,|\mathcal{\chi}|\}$
\STATE Do all the Hungarian steps for the cost matrix $\bold{P}'_2$.
\STATE Map the obtained solution to the assignment decision matrix $\boldsymbol{o}'$ which equals to $\boldsymbol{o}'=[\tau_{u_b}^{n}]$. \textbf{Then}
\STATE \textbf{For} $u=1:U_b$ in BS $b$
\STATE Check if \eqref{cp10} is satisfied or not
\STATE For all users ($u'\in\mathcal{U}''$) in BS $b$ which can not satisfy the constraint of \eqref{cp10}, \textbf{do}:
\STATE \textbf{For} $l_b^{\text{max}}$ times:
\STATE Introduce $\bold{P}''=\{P_{u',n}= |u'=1,2,\dots,|\mathcal{U}''|\,, n=1,2,\dots,|\mathcal{N}|\}$
\STATE Assign the minimum element of each row of $\boldsymbol{P}''$ which can satisfy \eqref{cp10} in the help of $\boldsymbol{o}'$.
\STATE \textbf{If} \eqref{dp8} is satisfied:
\STATE Assign the subcarrier to user $u$
\STATE \textbf{Else} \eqref{dp8} is not satisfied for subcarrier $n$:
\STATE Omit column $n$ in $\bold{P}''$
\STATE \textbf{End}
\end{algorithmic}
\end{algorithm*}
\vspace{-0.5em}\subsubsection{\textbf{Radio Power Allocation Sub-problem in Caching Phase}}\label{3d sp & s}
After solving the above sub-problems, the {corresponding solutions} are given to this sub-problem. As a result, the resource allocation problem in the caching phase is
\begin{align}\label{ro v tau sabet}
&
\minimize_{\boldsymbol{p}}\Bbb{E}_{\bold{H}(t)}\{\phi^{\text{Power}}(t)+\phi^{\text{BW-Radio}}(t)\}\\
&
\text{s.t. }~\nonumber
\eqref{cp5}, \eqref{cp6}, \eqref{cp7}, \eqref{cp10}.\nonumber
\end{align}
In \eqref{ro v tau sabet}, \eqref{cp10} makes the problem non-convex. {To  handle this, we apply DC and the details are summarized in Alg. 3. In the power allocation problem, SCA transforms the non-convex problem into a series of convex problems. Besides, the arithmetic-geometric mean (AGM) and DC approximations are two categories of SCA.
The basic idea behind the DC method is to exploit the first order Taylor approximation. For example, if the main function can be written as the difference of two concave functions $r=f-g$, the second term is approximated by a linear function.
 We first rewrite the data rate formula in its equivalent form as follows:}
\begin{align}\label{fog}
r_{b,u}^n=f_{b,u}^{n}-g_{b,u}^n, \,\ \forall n\in\mathcal{N}, \forall b\in\mathcal{B}, \forall u\in\mathcal{U},
\end{align}
where $f_{b,u}^n$ and $g_{b,u}^n$ are two concave functions which are formulated as
\begin{align}\label{f}
&
f_{b,u}^n=W\log\big(\tau_{b,u}^{n}p_{b,u}^{n}h_{b,u}^{n}+\sum\limits_{\scriptstyle i \ne u\atop
{\scriptstyle i\in \mathcal{U} \atop
\scriptstyle h_{b,i}^n\geq h_{b,u}^n}}\tau_{b,i}^{n}p_{b,i}^{n}h_{b,u}^{n}\\ \nonumber
&
+\sum\limits_{\scriptstyle j =0\atop
{\scriptstyle j \ne b}}^{B}\sum_{d\in \mathcal{U}_j} \tau_{j,d}^{n}p_{j,d}^{n}h_{j,u}^{n}+\sigma^2\big), \,\ \forall b\in\mathcal{B}, \forall u\in\mathcal{U},
\end{align}
and
\begin{align}\label{g}
&
g_{b,u}^n=W\log\Big(\sum\limits_{\scriptstyle i \ne u\atop
{\scriptstyle i\in \mathcal{U} \atop
\scriptstyle h_{b,i}^n\geq h_{b,u}^n}}\tau_{b,i}^{n}p_{b,i}^{n}h_{b,u}^{n}\\ \nonumber
&
+\sum\limits_{\scriptstyle j =0\atop
{\scriptstyle j \ne b}}^{B}\sum_{d\in \mathcal{U}_j} \tau_{j,d}^{n}p_{j,d}^{n}h_{j,u}^{n}+\sigma^2\Big), \,\ \forall b\in\mathcal{B}, \forall u\in\mathcal{U}.
\end{align}
Afterward, the upper bound of $g_{b,u}^{n}$ is obtained by applying Taylor series {for a fixed $\bold{p}^{(k-1)}$ from iteration $k-1\geq 0$ as follows}:
\begin{align}\label{gdc11}
g_{u,b}^{n,(k)}\leq g_{b,u}^{n,(k-1)}+\bigtriangledown g_{u,b}^{n,(k-1)}(\bold {p}^{(k)}-\bold{p}^{(k-1)}),
\end{align}
{where the right-hand side of \eqref{gdc11} is a concave function.} Here, $\bigtriangledown g_{b,u}^{n,(k-1)}=\frac{\partial g^{(k-1)}(\bold{p})}{\partial p_{b,u}^{n,(k-1)}},\, \forall b\in\mathcal{B},\,\forall u \in\mathcal{U}, \,n\in\mathcal{N}$, and is obtained by \eqref{grad1}.
\begin{figure*}[t]
\begin{align}\label{grad1}
\bigtriangledown g_{b,u}^{n,(k-1)}=\left\{
\begin{array}{ll}
0,
&
\hbox{if $i=u$,} \\
{\frac{W\sum\limits_{\scriptstyle i \ne u\atop
{\scriptstyle i \in \mathcal{U}\atop
\scriptstyle  h_{b,i}^n\geq h_{b,u}^n}}\tau_{b,i}^{n}h_{b,u}^{n}+\sum\limits_{\scriptstyle j =0\atop
{\scriptstyle j \ne b}}^{B}\sum_{d\in \mathcal{U}_j}\tau_{j,d}^{n}h_{j,u}^{n}}{\sum\limits_{\scriptstyle i \ne u\atop
{\scriptstyle i \in \mathcal{U}\atop
\scriptstyle  h_{b,i}^n\geq h_{b,u}^n}}\tau_{b,i}^{n}{p}_{b,i}^{n,(k-1)}h_{b,u}^{n}+\sum\limits_{\scriptstyle j =0\atop
{\scriptstyle j \ne b}}^{B}\sum_{d\in \mathcal{U}_j}\tau_{j,i}^{n}{p}_{j,i}^{n,(k-1)}h_{j,u_{b}}^{n}+\sigma^2}}, & \hbox{otherwise.}
\end{array}
\right.
\end{align}
\end{figure*}
As a consequence, by considering \eqref{fog}, \eqref{gdc11}, and \eqref{grad1}, the ergodic data rate approximation is established by
\begin{align}\label{dc1}
R_{b,u}^{n,\text{DC},(k)}=\Bbb{E}_{\bold{H}}\{f_{b,u}^{n,(k)}-g_{b,u}^{n,(k)}\}.
\end{align}
In this way, \eqref{cp10} can be recast as
\begin{align}\label{cp10dc}
&
\sum_{c=1}^{C}d_{c}\cdot\rho_{b.c}\cdot s_{c}\leq\sum_{u=1}^{U}\sum_{n=1}^{N}T\cdot R_{b,u}^{n,\text{DC},k}, \,\ \forall b\in\mathcal{B},
\end{align}
which is convex. 
 {In this case, for problem \eqref{ro v tau sabet}, starting with the feasible $\bold{p^{0}}$, the optimal $\bold{p^{k}}$ is determined by the following problem \cite{ngo2013joint}:}
\begin{align}\label{endsub22}
&
\minimize_{\boldsymbol{p}}\Bbb{E}_{\bold{H}}\{\phi^{\text{Power}}(t)+\phi^{\text{BW-Radio}}(t)\}\\
&
\text{s.t}~\nonumber 
\eqref{cp5}, \eqref{cp6}, \eqref{cp7}, \eqref{cp10dc}.\nonumber
\end{align}
{Note that $\bold{p}^{k-1}$ is found from  iteration $k-1$. Moreover, the value of $\bold{p}^{k}$ is used for finding the value of $\bold{p}^{k+1}$}. Also, \eqref{endsub22} is a convex problem and can be solved by MATLAB tools like CVX. The pseudo code of the proposed SCA algorithm with the D.C. approximation method is summarized in Alg. 5.
\begin{algorithm}[tp]\label{SCA}
\caption{The proposed SCA algorithm with the D.C. approximation method for solving \eqref{ro v tau sabet}}
\begin{algorithmic}[1]
\STATE With the given $\tau_{b,u}^{n}(t)$ and $\rho_{b,c}(t)$
\STATE~~\textbf{repeat}
\STATE~~~~ Calculate $R_{b,u}^{n,\text{DC},k}$ by using \eqref{dc1}.
\STATE~~~~\textbf{Find} $\bold{p}$ by solving \eqref{endsub22}
\STATE~~~~ Update $k$ to $k+1$
\STATE~~\textbf{Until} $\bold{p}$ is converged.
\STATE \textbf{Output}: The power allocation
\end{algorithmic}
\end{algorithm}
\vspace{-0.5em}\section{\textbf{Problem Formulation and Solution of the Delivery Phase}}\label{pfsd}
\subsection{Objective Function of The Delivery Phase}
\paragraph*{\textbf{\textit{Total cost of the delivery phase}}}
Let $y_{i,b,c}(t)\in \{0,1\}$  be the cooperative fetching indicator of content $c$ from BS $i$ by BS $b$ and
$z_{b,c}(t)\in\{0,1\}$ be the binary indicator of determining whether content $c$ is downloaded from the remote CP or not. If
$z_{b,c}(t)=1$, content $c$ is fetched from the backhaul link and $z_{b,c}(t)=0$ otherwise. The third part of our considered cost is the bandwidth consumption of the optical fiber links in the whole  network which is given by
\begin{align}
\nonumber
&\phi^{\text{Link-BW}}(t)=\sum_{b=0}^{B}\sum_{c=1}^C\big(\sum\limits_{\scriptstyle i = 0\hfill\atop\scriptstyle b \ne i\hfill}^{B}C^{ \text{FH}}\cdot y_{i,b,c}(t)\cdot r_{i,b,c}^{\text{FH}}(t) +\nonumber\\
&
C^{\text {BH}}\cdot z_{b,c}(t)\cdot r_{b,c}^{\text{BH}}(t)\big), \,\ \forall b\in\mathcal{B},
\end{align}
where $C^{\text{FH}}$ and $C^{\text{BH}}$ are, respectively, the delivery cost of CCHC and CMC, and
 their units are $\frac{\text{price}}{\text{bps}}$. Moreover, $r_{i,b,c}^{\text{FH}}(t)$ is the data rate of content $c$ in the fronthaul link from BS $i$ to BS $b$ $(i\neq b)$ at time slot $ t $, and  $r_{b,c}^{\text{BH}}(t)$ is the allocated data rate of the backhaul link in BS $b$ to deliver  content $c$ at time slot $ t $.
 Note that since the cost of the fronthaul links is less than the cost of
the backhaul links, we have 
$C^{\text {FH}}  < C^{\text {BH}}$.
Finally, the total cost of the network at time slot $ t $ is calculated by 
\begin{align}
\phi^{\text{Total}}(t)=\phi^{\text{Power}}(t)+\phi^{\text{BW-Radio}}(t)+\phi^{\text{Link-BW}}(t).
\end{align}
\paragraph*{\textbf{\textit{Delivery Constraints}}}
$x_{b,c}(t)\in \{0,1\}$ is considered as a binary variable to show the occurrence of CHC for content $c$ in BS $b$ at time slot $ t $. Thus, if
$x_{b,c}(t)$ is 1, content $c$ is transmitted from the storage of BS $b$, and $x_{b,c}(t)=0$ otherwise. As mentioned before,
the content delivery process finishes when all the requested contents are delivered. 
Moreover, each content can be fetched by
CHC, CCHC, CMC. Therefore, the following constraint denotes that the content delivery process in each BS $b$ for each content $c$ should be performed by {only} one of the provisioning cases:}
\begin{align} \label{dp11}
x_{b,c}(t)+\sum\limits_{\scriptstyle i = 0\hfill\atop\scriptstyle i \ne b\hfill}^{B}y_{i,b,c}(t)+z_{b,c}(t)\leq 1, \,\ \forall b \in\mathcal{B}, c\in\mathcal{C}.
\end{align}
If $\rho_{b,c}(t')$ is 0, $x_{b,c}(t)$ cannot be 1. In other words, CHC cannot occur if the requested content $c$ does not exist in the storage of associated BS $b$. Therefore, {it yields} the following constraint:
\begin{align}\label{dp3}
x_{b,c}(t)\leq\rho_{b,c}, \,\ \forall b\in\mathcal{B}, c\in\mathcal{C}.
\end{align}
On the other hand, $y_{i,b,c}(t)$ can be 1, if content $c$ is cached in the storage of BS $i$. In this case, {the} CCHC occurrence {of} content $c$ in BS $b$ depends on the existence of content $c$ in the storage of BS $i$, and hence, we have
\begin{align}\label{dp4}
y_{i,b,c}(t)\leq\rho_{i,c}, \,\ \forall i,b\in\mathcal{B}, i\neq b, c\in\mathcal{C}.
\end{align}
\paragraph*{{\textbf{\textit{Traffic Constraint}}}}
Note that more than one user in BS $b$ could request for content $c$. To deliver content $c$ from BS $i$ to BS $b$ in CCHC case, the dedicated data rate of fronthaul link from BS $i$ to BS $b$ should be {larger} than the minimum access rate of requesting users. Therefore, we introduce the following constraint:
\begin{align}\label{dp1}
&
\min _{u , \delta_{u,c}(t)=1}
\{\sum_{n=1}^{N}r_{b,u}^{n}(t)\}y_{i,b,c}(t)\leq y_{i,b,c}(t)r_{i,b,c}^{\text{FH}}(t), \,\ \nonumber\\
&
\forall i, b\in\mathcal{B}, i\neq b, c\in\mathcal{C}.
\end{align}
{where $\min$ {denotes} the minimum value.} {Following} the same {argument} as above, for the fetching from the backhaul link to BS $b$ in CMC case, the following constraint is included:
\begin{align}\label{dp2}
\min _{u , \delta_{u,c}(t)=1}\{\sum_{n=1}^{N}r_{b,u}^{n}(t)\}z_{b,c}(t)\leq z_{b,c}(t)r_{b,c}^{\text{BH}}(t), \,\  \forall b\in\mathcal{B}, c\in\mathcal{C}.
\end{align}
\paragraph*{\textbf{\textit{Content Delivery Time Constraint}}}
{We suppose each requested content is enforced to be delivered in each short time slot which is $T$ seconds to avoid the latency}. Moreover, the latency of the fronthaul and backhaul links latency is assumed to be ultra-low \cite{tran2017cooperative}. 
{Hence, the next constraint controls the delivery time for each content and each user as follows:
\begin{align}\label{dp10}
&
\sum_{n=1}^{N}\tau_{b,u}^{n}(t)
\delta_{u,c}(t)s_c
\leq \delta_{u,c}(t)(\sum_{n=1}^{N}\tau_{b,u}^{n}(t))(\sum_{n=1}^{N}r^{n}_{b,u}(t)\cdot T),\\
&
\,\ \forall b\in\mathcal{B}, c\in\mathcal{C}, u\in\mathcal{U},\nonumber
\end{align} 
}
The above constraint shows that each content $c$ with the size of $s_c$ should be delivered in $T$ seconds to user $u$ in BS $b$.
\paragraph*{\textbf{\textit{Maximum Capacity for Fronthaul Fiber Link Constraint}}}
The allocated fronthaul data rate for all contents transmitted from BS $i$ to BS $b$ should not exceed the finite maximum data rate of the mentioned fronthaul link. Therefore, we have
\begin{align}\label{dp12}
\sum_{c=1}^{C}r_{i,b,c}^{\text{FH}}(t)\leq R_{i,b}^{\text{max},\text{FH}}, \,\ \forall i, b\in\mathcal{B}, i\neq b,
\end{align}
where $R_{i,b}^{\text{max},\text{FH}}$ denotes the maximum data rate of the fronthaul link between BS $i$ and BS $b$.
\vspace{-0.5em}\subsection{\textbf{Delivery Phase}}\label{g delivershode}
As it is shown in Fig. \ref{framework}, the delivery phase is composed of several time periods of $T$ seconds. The length of $T$
is assumed to be much smaller than the whole considered time for the delivery phase. In this work, for each small time slot, the
channel fading is assumed to be fixed and in every of them it changes independently from the last time slot. \\
In the delivery phase, the BSs serve contents by their own cache storages or via the fronthaul links or via the backhaul
link. According to the proposed system model, the problem formulation for the delivery phase is
\begin{align}
\label{delivery}
&
\minimize_{\boldsymbol{r^{\text{BH}}, r^{\text{FH}}, p, \tau, x, y, z}}\phi^{\text{tot}}\\
&
\text{s.t. }~~~~\eqref{dp5}-\eqref{dp9}, \eqref{dp11}-\eqref{dp12},\nonumber
\end{align}
where $\bold{r}^{\text{FH}}=[r_{i,b,c}^{\text{FH}}(t)],\,\bold{r}^{\text{BH}}=[r_{b,c}^{\text{BH}}(t)],\,\bold{x}(t)=[x_{b,c}(t)],\,\bold{y}(t)=[y_{i,b,c}(t)]$, and $\bold{z}(t)=[z_{b,c}(t)]$. The problem is MINLP and non-convex. Hence, we apply ASM.
Thus, the problem is divided into three sub-problems. In this regard, the first one is about allocating subcarrier, the second
one is about distinguishing content provisioning cases and the last one is about resource allocating. The ASM for the delivery case is summarized in Alg. 4. {Also, the overview of proposed algorithms for the delivery phase is shown in Fig. 5. The initial point of the caching phase can be found on Appendix B.}
\begin{figure}\label{fchdel}
	\centering
	\includegraphics[width=0.4\textwidth]{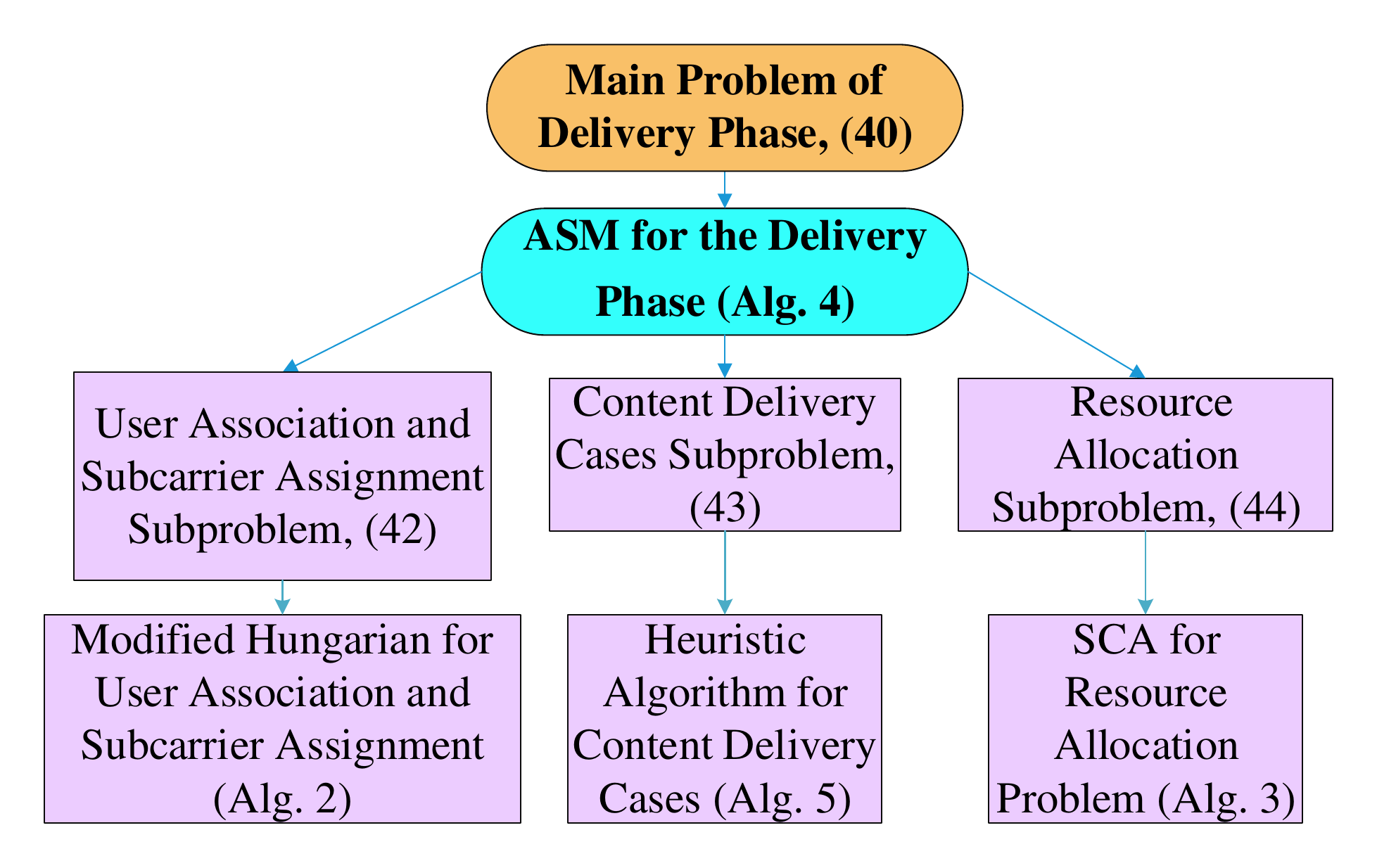}
	\caption{The overview of the proposed algorithm in the delivery phase.}
\end{figure}
\vspace{-0.75em}\subsubsection{\textbf{User Association and Subcarrier Assignment Sub-problem in Delivery Phase}}
The considered subcarrier assignment non-convex problem is given by
\begin{align}\label{sub3}
&
\minimize_{\boldsymbol{\tau}}\phi^{\text{tot}}\\
&
\text{s.t. }~~~~\eqref{dp5}, \eqref{dp8}-\eqref{dp9}, \eqref{dp1}-\eqref{dp10}.\nonumber~~
\end{align}
 The solution for this problem is similar to the algorithm which is described in Section \ref{2nd sp & s} and summarized in Alg. 3.
 \begin{algorithm}
 \caption{{The ASM for the delivery phase}}
 \begin{algorithmic}[1]\label{ASMCACH1}
 \STATE For $k_1=1$ to $K_1$ which is the maximum number of iteration, and $0<\Upsilon<<1$ is the tolerance (or accuracy):
 \STATE \textbf{repeat}
 \STATE~~Find $\boldsymbol{\tau}^{(k_1)}$ by solving \eqref{sub3} via using Hungrain Algorithm which is stated in Alg. 2.
 \STATE~~Find $\boldsymbol{x}^{(k_1)}, \boldsymbol{y}^{(k_1)}, \boldsymbol{z}^{(k_1)}$ by solving \eqref{sub-problem1} via employing heuristic algorithm.
 \STATE~~Find $\bold{p}^{(k_1)}$, $\bold{r}^{\text{FH},(k_1)}$ and $\bold{r}^{\text{BH},(k_1)}$  by solving \eqref{sub21} via applying SCA-DC Approximation which is presented in Alg. 3.
 \STATE  $k_1=k_1+1$
 \STATE \textbf{Until} $|\phi^{\text{tot},(k_1+1)}-\phi^{\text{tot},{(k_1)}}|$$\le$$\epsilon$.
 \STATE \textbf{Output} $\bold{\tau},\,\bold{x},\,\boldsymbol{y},\,\bold{z},\,\bold{p},\,\bold{r}^{\text{FH}},\,\bold{r}^{\text{BH}}$.
 \end{algorithmic}
 \end{algorithm}
\subsubsection{\textbf{Content Delivery Cases Sub-problem}}\label{cd}
{After assigning the subcarriers to the users, we need to determine how to deliver the content $c$ (cache hit, cooperative caching or cache miss) to the users of each BS $b$.} The optimization
problem for this problem is
\begin{align}\label{sub-problem1}
&
\minimize_{\boldsymbol{x, y, z}}\phi^{\text{tot}}\\
&
\text{s.t}~~~~ \eqref{dp11}-\eqref{dp2},\nonumber
\end{align}
which is ILP problem. {To deal with this problem,} we propose a novel heuristic algorithm, which is summarized in {Alg. 5. With a given} $\boldsymbol{\tau}$, we determine all the users who associate to BS $b$ and we put the user's index number in vector $\bold{u}_b$. In the next step, we search the request matrix $\boldsymbol{\delta}$ for all users in vector $\boldsymbol{u}_b$ to see which contents are requested in BS $b$ and then, we put the requested contents' index numbers in vector$\boldsymbol{c}'_b$.
 For each content $c'\in\boldsymbol{c}'_b$ , we check the content placement matrix $\boldsymbol{\rho}$.
If $\rho_{b,c'}=0, \,\ \forall b \in\mathcal{B}$, set $z_{b,c'}=1$, $x_{b,c'}=0$ and $y_{i,b,c'}=0$. Else if
$\rho_b^{c'}=1$ then, set $x_{b}^{c'}=1$ and $y_{i,b}^{c'}=0$, $z_{b}^{c'}=0$. If $\rho_i^{c'}=1,
\,\ \forall i\in\mathcal{B}, i\neq b$, for each content $c'$ fetch the index number of BS $i$ and put it in vector
$\boldsymbol{\rho}^{c'}$. Then for each $i'\in\boldsymbol{\rho}^{c'}$, we check which one is resulted in the minimum
objective function value in optimization problem and put the BS number in vector $\boldsymbol{i}_{'c'}$ (If there is more
than one BS number in $\boldsymbol{i}_{'c'}$, choose one of them randomly.)\\
We determine the minimum BS assignment for each content $c'$. Then, for all contents which have the identical number of
selected BS check whether \eqref{dp12} is satisfied or not, and if \eqref{dp12} is satisfied, put $y_{i',b}^{c'}=1, \,\
\forall i' \in\boldsymbol{i}_{'c'}$, $x_{b,c'}=0$ and $z_{b,c'}=0$, otherwise, calculate $r_{i',b,c'}^{\text{FH}}$ for
each content $c'$ then put it in vector $\boldsymbol{R}^{',\text{FH},c'}$. Then, sort the elements of $\boldsymbol{R}_{c'}^{',\text{FH}}$, increasingly. After all these steps, from the
first (the smallest) element of $\boldsymbol{R}_{c'}^{',\text{FH}}$ set $z_{b,c'}=1$, $x_{b,c'}=0$ and $y_{i',b,c'}=0$, until
\eqref{dp12} get satisfied then for the remain contents index number in $\boldsymbol{R}^{',\text{FH},c'}$ set
$y_{i',b,c'}=1, \,\ \forall i' \in\boldsymbol{i}_{'c'}$, $x_{b,c'}=0$ and $z_{b,c'}=0$.
\begin{algorithm}[tp]
\caption{{Heuristic algorithm for content delivery: performer at each time slot $ t $}}
\begin{algorithmic}[1]
\STATE \textbf{Input} $\boldsymbol{\rho}$, $\boldsymbol{\tau}$, $\boldsymbol{\delta}$
\STATE \textbf{Output} $\bold{x}$, $\bold{y}$ and $\bold{z}$,
\STATE~~\textbf{For} $b=1$ to $b=B$  \textbf{do}
\STATE~~~~~~Find the set of users that are in BS $b$ by scanning $\boldsymbol{\tau}$ and put the users number in vector $\boldsymbol{u}_b$,
\STATE~~~~~~ Find requested contents by scanning the requested matrix $\boldsymbol{\delta}$, for all users in $\boldsymbol{u}_b$ and put them in vector $\boldsymbol{c'}_b$
\STATE~~~~~~~~\textbf{End}
\STATE~~~~~~~~\textbf{For} $c'\in \boldsymbol{c}'_{b}$ check matrix $\boldsymbol{\rho}$ find ${\rho}_{b',c'}(t-1)=1, \,\ \forall b'\in\mathcal{B}$ and put the BS index in vector $\boldsymbol{\rho}_{c'}$. \textbf{IF} $\boldsymbol{\rho}_{c'}=\{\}$, then:
\STATE~~~~~~~~~~~~~Put $z_{b',c'}=1$, $y_{i,b',c'}=0$ and $x_{b',c'}=0$
\STATE~~~~~~~~~~~~~ \textbf{Else if}
\STATE~~~~~~~~~~~~~$\rho_{b',c'}=1$ and $b=b'$ then:
\STATE~~~~~~~~~~~~~~~Put $x_{b',c'}=1$, $y_{i,b',c'}=0$ and $z_{b',c'}=0,\,\,\forall i\in\mathcal{B}, i\neq b'$
\STATE~~~~~~~~~~~~~\textbf{Else}, $\rho_{b',c'}=1$ and $b\neq b'$ then:
\STATE~~~~~~~~~~~~~Check the objective for the minimum assignment of BS $i \in\boldsymbol{\rho}_{c'}$ and fetch the BS number of minimum assignment and put it in $\boldsymbol{i}_{'c'}$. (if $|\boldsymbol{i}_{'c'}|>1$, choose one of the elements randomly.)
\STATE~~~~~~~~~~~~~ For all contents where $\boldsymbol{i}_{'c'}$ is identical, check if \eqref{dp12} is satisfied or not. \textbf{If} \eqref{dp12} is satisfied:
\STATE~~~~~~~~~~~~~~~Put $y_{i',b,c'}=1$, $x_{b,c'}=0$ and $z_{b,c'}=0, \,\ \forall i'\in\boldsymbol{i}_{'c'}$
\STATE~~~~~~~~~~~~~~~\textbf{Else}
\STATE~~~~~~~~~~~~~~~For all contents which need to be delivered from BS $i'$, \textbf{do}:
\STATE~~~~~~~~~~~~~~~~~~ Calculate $r_{i',b,c'}^{\text{FH}}$ and sort them increasingly in vector $\boldsymbol{R}_{c'}^{'\text{FH}}$
\STATE~~~~~~~~~~~~~~~~~~ From the first element of $\boldsymbol{R}^{'\text{FH},c'}$, for each content of $c'$ put $z_{b,c'}=1$, $y_{i',b,c'}=0$ and $x_{b,c'}=0$, \textbf{until} \eqref{dp12} is satisfied. Then, for the remain contents put $y_{i',b,c'}=1$, $x_{b,c'}=0$ and $z_{b,c'}=0, \,\ \forall i'\in\boldsymbol{i}_{'c'}$.
\STATE~~~~~~~~~~\textbf{End}
\STATE~~~~~~~~~In each step of this algorithm total capacity of network is exceeded, the requested content is rejected for the current time slot.
\end{algorithmic}
\label{heuristic}
\end{algorithm}
\vspace{-0.75em}\subsubsection{\textbf{Radio Power and Optical Links Resource Allocation Sub-problem in Delivery Phase}}\label{RRA}
In the last step of solving the delivery phase problem, the resources should be allocated to the users. The proposed algorithm for this sub-problem is summarized in Alg. 4. Note that, in this sub-problem $x_{b,c}$ , $y_{i,b,c}$ , $z_{b,c}$ and $\tau_{b,u}^{n}$ are fixed and evaluated in the last sub-problems. In the proposed resource allocation scheme, we aim to find $p_{b,u}^{n}$ , $r_{i,b,c}^{\text{FH}}$, and $r_{b,c}^{\text{BH}}$ by solving the following problem:
\begin{align}
&
\label{sub21}
\minimize_{\boldsymbol{r^{\text{BH}}, r^{\text{FH}}, p}}\phi^{\text{tot}}\\
\text{s.t}~~~~\eqref{dp5}, \eqref{dp6}, \eqref{dp7}, \eqref{dp1}-\eqref{dp12}.\nonumber 
\end{align}
Obviously, \eqref{sub21} is non-convex problem because of non-convexity of \eqref{dp1}, \eqref{dp2}, and \eqref{dp10}. {To deal with this problem, SCA is applied.} From \eqref{fog}, \eqref{f}, \eqref{g}, and \eqref{grad1}, $r^{n,\text{DC},(k)}_{b,u}$ is introduced to approximate the data rate by
\begin{align}\label{dc2}
r^{n,\text{DC},(k)}_{b,u}=f_{b,u}^{n,(k)}-g_{b,u}^{n,(k)}.
\end{align}
Thereby, \eqref{dp10} can be reformulated to:
\begin{align}\label{dcdp10}
&
\sum_{n=1}^{N}\delta_{u,c}\tau_{b,u}^{n}s_c\leq\delta_{u,c} (\sum_{n=1}^{N}\tau_{b,u}^{n}T)(\sum_{n=1}^{N}r^{n,\text{DC},(k)}_{b,u}), \,\ &\forall c\in\mathcal{C},\nonumber\\
&
 u\in\mathcal{U}, b\in\mathcal{B}.
\end{align}
Additionally, to {deal with the non-convexity of} \eqref{dp1} and \eqref{dp2}, the upper bound of $f_{b,u}^n$ is obtained by Taylor series to:
\begin{align}\label{fdc11}
f_{b,u}^{n,(k)}\leq f_{b,u}^{n,(k-1)}+\bigtriangledown f_{b,u}^{n,k-1}(\bold{p}^{(k)}-{\bold{p}}^{(k-1)}),
\end{align}
where $\bigtriangledown f_{b,u}^{n,(k-1)}(\bold{p})=\frac{\partial f^{(k-1)}(\bold{p})}{\partial p_{b,u}^{n,(k-1)}},\, \forall b\in\mathcal{B},\,\forall u \in\mathcal{U}, \,n\in\mathcal{N}$ and is given by \eqref{fgrad}.
\begin{figure*}[t]
\begin{align}\label{fgrad}
\bigtriangledown f_{b,u}^{n,(k-1)}(\bold{p})=\left\{
\begin{array}{ll}
 0,
&
\hbox{if $i=u$,} \\
{\frac{\sum\limits_{\scriptstyle i \ne u\atop
{\scriptstyle i \in \mathcal{U}\atop
\scriptstyle  h_{b,i}^n\geq h_{b,u}^n}}h_{b,u}^{n}\tau_{b,i}^{n}+\sum\limits_{\scriptstyle j =0\atop
{\scriptstyle j \ne b}}^{B}{\sum_{j\in \mathcal{U}}\tau_{j,d}^{n}h_{j,u}^{n}}+\tau^{n}_{b,u}h_{b,u}^{n}}{\sum\limits_{\scriptstyle i \ne u\atop
{\scriptstyle i \in \mathcal{U}\atop
\scriptstyle  h_{b,i}^n\geq h_{b,u}^n}}\tau_{b,i}^{n}{p}_{b,i}^{n,(k-1)}h_{b,u}^{n}+\sum\limits_{\scriptstyle j =0\atop
{\scriptstyle j \ne b}}^{B}{\sum_{i\in \mathcal{U}} \tau_{j,i}^{n}{p}_{j,i}^{n,(k-1)}h_{j,u}^{n}}+\sigma^2+\tau^{n}_{b,u}h_{b,u}^{n}p_{b,u}^{n,(k-1)}}}, & \hbox{otherwise}.
\end{array}
\right.
\end{align}
\end{figure*}
Hence, from \eqref{fog}, \eqref{f}, \eqref{g}, \eqref{fdc11}, and \eqref{fgrad}, we introduce $r^{'n,\text{DC},(k)}_{b,u}$ as
\begin{align}\label{rprim}
r_{b,u}^{'n,\text{DC},(k)}=g_{b,u}^{n,(k)}-f_{b,u}^{n,(k)}.
\end{align}
Finally, \eqref{dp1} and \eqref{dp2} can be written as
\begin{align}\label{dcdp2}
\min _{u , \delta_{u,c}=1}\{\sum_{n=1}^{N}r^{'n,\text{DC},(k)}_{b,u}\}y_{i,b,c}\leq  y_{i,b,c}r_{i,b,c}^{\text{FH}}, \,\ \forall i, b\in\mathcal{B}, i\neq b, c\in\mathcal{C}
\end{align}
\begin{align}\label{dcdp3}
\min _{u , \delta_{u,c}=1}\{\sum_{n=1}^{N}r^{'n,\text{DC},(k)}_{b,u}\}z_{b,c}\leq z_{b,c}r_{b,c}^{\text{BH}}, \,\  \forall b\in\mathcal{B}, c\in\mathcal{C},
\end{align}
respectively. Since all the constraints become convex, \eqref{sub21} can be written as
\begin{align}\label{sub2}
&\min_{\boldsymbol{r^{\text{BH}}, r^{\text{FH}}, p}}\phi^{\text{tot}}\\
\text{s.t}~\nonumber 
&
\eqref{dp5}, \eqref{dp6}, \eqref{dp7}, \eqref{dcdp10}, \eqref{dp12}, \eqref{dcdp2}, \eqref{dcdp3}. \nonumber
\end{align}
which is a convex problem and can be solved by CVX tools in MATLAB software.
{
\section{CONVERGENCE AND COMPLEXITY Analysis}   \label{The Convergence of iterative algorithm and DC approximation}
In this section, we discuss the convergence and complexity of the adopted solution. The convergence of the ASM algorithm and SCA DC is proven in the Appendix C and Appendix D, respectively.
}

\paragraph*{\underline{\textit{\textbf{The Monotonicity Proof of Heuristic Algorithm}}}}
The delivery phase initial point is feasible and is obtained by the initial point optimization problem for the delivery phase. Moreover, we obtain $x_{b,c}^{(q)}$, $y_{i,b,c}^{(q)}$, and $z_{b,c}^{(q)}$ from the delivery cases sub-problem. If $x_{b,c}^{(q)}=1$, $y_{i,b,c}^{(q)}=0$, and $z_{b,c}^{(q)}=0$, we have
\begin{align}
&
\phi(\bold{r}^{\text{BH},(q)}, \bold{r}^{\text{FH},(q)}, \bold{p}^{(q)}, \bold{x}^{(q)}, \bold{y}^{(q)}, \bold{z}^{(q)}, \boldsymbol{\tau}(t)^{(q+1)})
\nonumber \\
&
\geq\phi(\bold{r}^{\text{BH},(q)}, \bold{r}^{\text{FH},(q)}, \bold{p}^{(q)}, \bold{x}^{(q+1)}, \bold{y}^{(q+1)}, \bold{z}^{(q+1)}, \boldsymbol{\tau}^{(q+1)}).
\end{align}
Furthermore, if $x_{b,c}^{(q)}=0$, $y_{i,b,c}^{(q)}=1$, and $z_{b,c}^{(q)}=0$, the objective function decreases in each iteration, based on line 13 in Alg. 7. Thus, we have
\begin{align}
&
\phi(\bold{r}^{\text{BH},(q)}, \bold{r}^{\text{FH},(q)}, \bold{p}^{(q)}, \bold{x}^{(q)}, \bold{y}^{(q)}, \bold{z}^{(q)}, \boldsymbol{\tau}^{(q+1)})
\nonumber \\
&
\geq\phi(\bold{r}^{\text{BH},(q)}, \bold{r}^{\text{FH},(q)}, \bold{p}^{(q)}, \bold{x}^{(q+1)}, \bold{y}^{(q+1)}, \bold{z}^{(q+1)}, \boldsymbol{\tau}^{(q+1)}).
\end{align}
Otherwise, ($x_{b,c}^{(q)}=0$, $y_{i,b,c}^{(q)}=0$, and $z_{b,c}^{(q)}=1$), the objective function value drops in each iteration based on lines 7 and 19 in Alg. 7, we have
\begin{align}
&
\phi(\bold{r}^{\text{BH},(q)}, \bold{r}^{\text{FH},(q)}, \bold{p}^{(q)}, \bold{x}^{(q)}, \bold{y}^{(q)}, \bold{z}^{(q)}, \boldsymbol{\tau}^{(q+1)})
\nonumber \\
&
\geq\phi(\bold{r}^{\text{BH},(q)}, \bold{r}^{\text{FH},(q)}, \bold{p}^{(q)}, \bold{x}^{(q+1)}, \bold{y}^{(q+1)}, \bold{z}^{(q+1)}, \boldsymbol{\tau}^{(q+1)}).
\end{align}
In this regard, the monotonicity of ASM is proofed.
\begin{figure}[t]
\centering		
\subfigure[]{ \includegraphics[width=.4\textwidth]{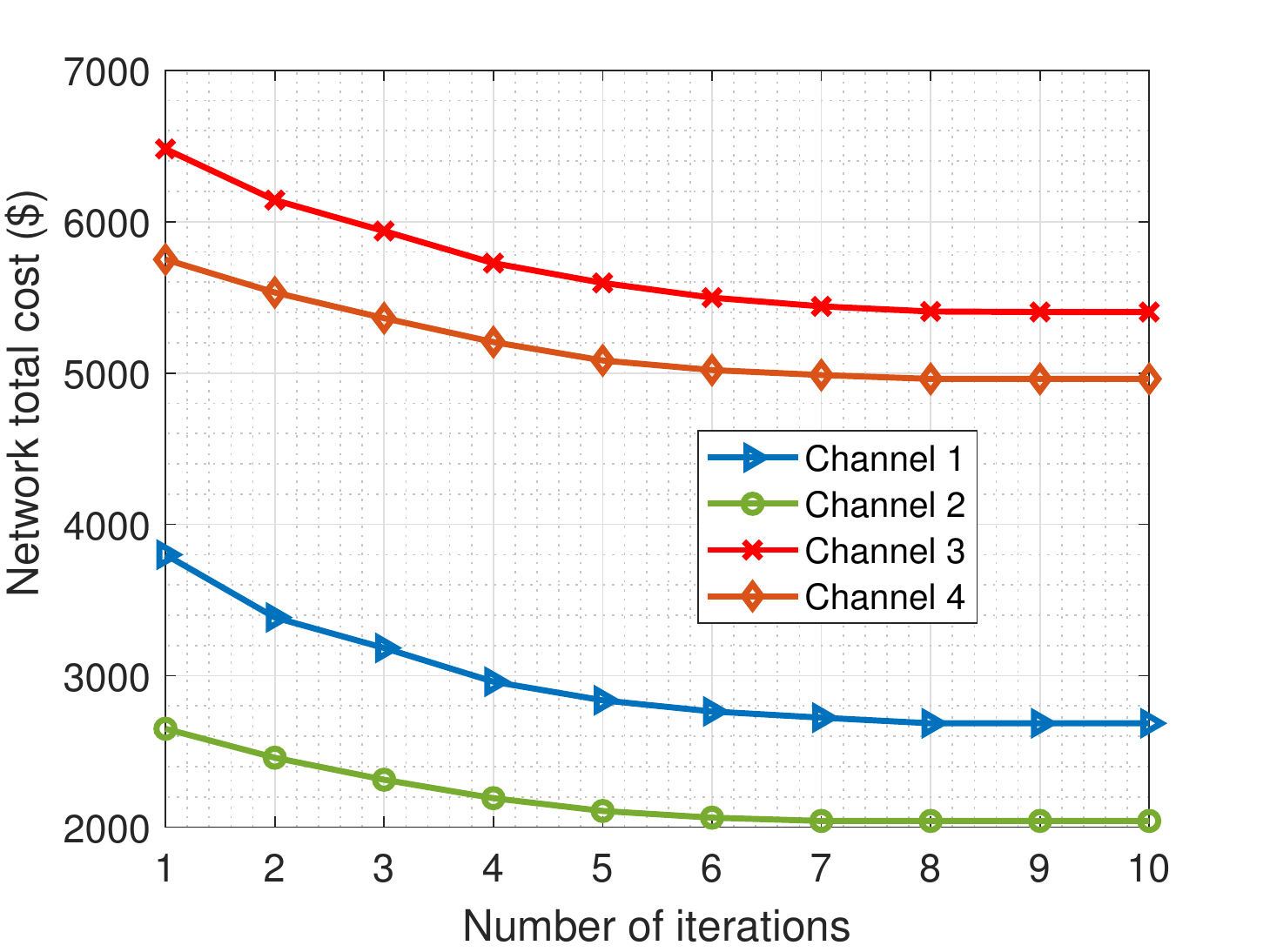}}
\subfigure[]{\includegraphics[width=.4\textwidth]{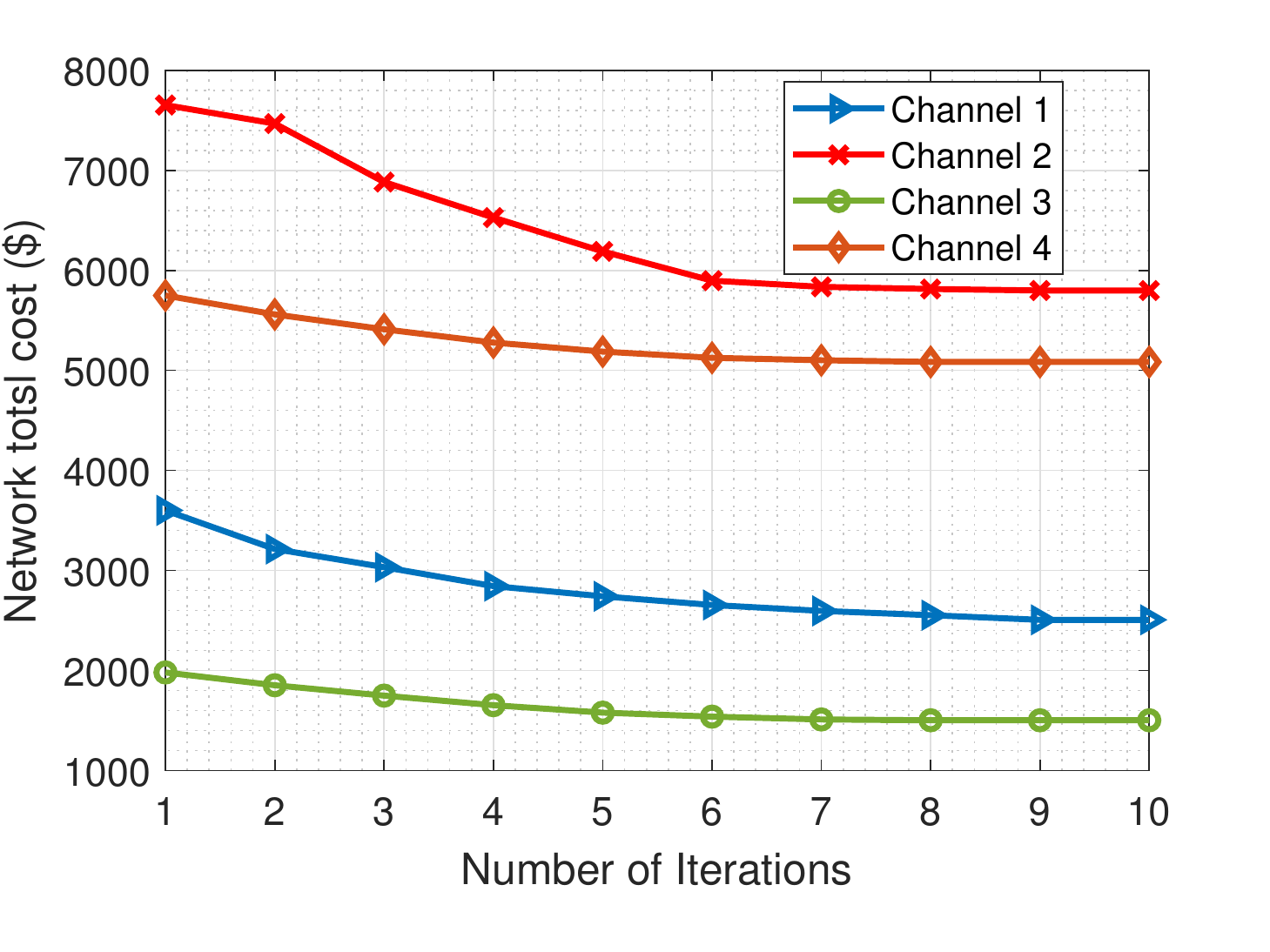}}
\caption{ An example of convergences of proposed solution: (a): Converges of SCA Algorithm in delivery phase. (b): Converges of ASM Algorithm in the delivery phase.}
\label{asmsca}
\end{figure}
	\begin{figure*}
\begin{align}
{\label{M2Constraint}
M_2=(B\times [B-1]\times C)+(B\times C)+B+2\cdot(U\times B\times N)+(U\times B\times C)+B^2=\mathcal{O}\big(B\times C[B+C]\big).
}
\end{align}
\hrule
\end{figure*}
\subsection{{Complexity Comparison between Our Proposed Network and NC-OMA}}
The computational complexity can be found in Appendix E. Also, the complexity for each employed method for the cache phase and the delivery phase are summarized in Table \ref{com} and Table \ref{NC-OMA}, respectively. {To compare the cost reduction of our proposed network, we consider different number of users and SBSs in the network.
All of these parameters are
compared to the NC-OMA scheme from the complexity and cost reduction point of view in Table \ref{Trade_Off}.
 In this case for NC-OMA, some
constraints are omitted which are \eqref{dp1}, \eqref{dp4}, \eqref{dp7}, and \eqref{dp12} in the caching phase and the
delivery phase which means that the NC-OMA network is less complex compared to our considered network. The complexity of NC-OMA is
summarized in Table \ref{NC-OMA} where
}
$ M_4= B\times C+B, $
and
{
\begin{align}
\nonumber
&
M_5= M_6 = B\times C +B +U\times B\times N+ B\times C\times U+ B
\\&\approxeq\mathcal{O}\left(B\times C\times U\right).
\end{align}
}
\begin{table*}[!ht]
	\centering
	\caption{Complexity and cost reduction trade-off between CO-NOMA and NC-OMA}
	\label{Trade_Off}
	\begin{tabular}{|l||l|l|l|l|l|l|l|l|l|}
		\hline
		\begin{tabular}[c]{@{}l@{}}CO-NOMA over CO-OMA\end{tabular}      & \begin{tabular}[c]{@{}l@{}}U=15\\ B=4\end{tabular} & \begin{tabular}[c]{@{}l@{}}U=15\\ B=8\end{tabular} & \begin{tabular}[c]{@{}l@{}}U=35\\ B=4\end{tabular} & \multicolumn{1}{c|}{\begin{tabular}[c]{@{}c@{}}U=35\\ B=8\end{tabular}} & \begin{tabular}[c]{@{}l@{}}U=55\\ B=4\end{tabular} & \begin{tabular}[c]{@{}l@{}}U=55\\ B=8\end{tabular} & \begin{tabular}[c]{@{}l@{}}U=85\\ B=4\end{tabular} & \begin{tabular}[c]{@{}l@{}}U=85\\ B=8\end{tabular} & Average value \\ \hline\hline
		\begin{tabular}[c]{@{}l@{}}Complexity metric (\%)\end{tabular}  &         \multicolumn{1}{c|}{0.3433}                                            &               \multicolumn{1}{c|}{0.6866}                                      &  \multicolumn{1}{c|}{1.8692}                                                    &                    \multicolumn{1}{c|}{3.7384 }                                                     &   \multicolumn{1}{c|}{4.6158 }                                                 &                  \multicolumn{1}{c|}{ 9.2316}                                  &     \multicolumn{1}{c|}{ 4.7059}                                               &                           \multicolumn{1}{c|}{9.4118}                            &  \multicolumn{1}{c|}{ 4.3362}         \\ \hline\hline
		\begin{tabular}[c]{@{}l@{}}Cost metric (\%)\end{tabular} &        \multicolumn{1}{c|}{36}                                             &          \multicolumn{1}{c|}{41}                                           & \multicolumn{1}{c|}{30}                            &    \multicolumn{1}{c|}{51}                                                                      &            \multicolumn{1}{c|}{28}                                         &                   \multicolumn{1}{c|}{48}                                  & \multicolumn{1}{c|}{21}                            &                            \multicolumn{1}{c|}{46}                         &    \multicolumn{1}{c|}{37.625}      \\ \hline
	\end{tabular}
\end{table*}
{Moreover, Table \ref{NC-OMA} shows the performance and complexity of our proposed scheme compared to NC-OMA. {By taken average, on the achieved gain values of Table \ref{Trade_Off}, we   obtain that with an acceptable increase in the complexity, 
		the cost reduction is considerable, as numerically $37.625  $\% in general.}
}
\begin{table}
\centering
\caption{Complexity order of the proposed solution for caching and delivery phases in the considered CO-NOMA}
\label{com}
\begin{tabular}{|L{1cm}|L{2.7cm}|L{1.1cm}|L{1.1cm}|L{1cm}|}
\hline
\cellcolor{green}{Phase}&	\cellcolor{green}{User association and subcarrier assignment sub-problem}  & \cellcolor{green}{Content placement sub-problem}&\cellcolor{green}{Radio resource allocation sub-problem}& \cellcolor{green}{Delivery cases sub-problem}\\	
\hline
\cellcolor{orange}{Caching phase} &$\mathcal{O}(|\text{max}\{|U|,|B|\}|)^{3}+\mathcal{O}(|\text{max}\{|U|,|N|\}|)^{3}
+\mathcal{O}(U\times N)$& $\frac{\log\left(\frac{M_3}{t^0\varrho}\right)}{\log(\vartheta)}$&$\frac{\log\left(\frac{M_1}{t^0\varrho}\right)}{\log(\vartheta)}$&$-$\\
\hline	
\cellcolor{orange}{Delivery phase} & $\mathcal{O}(|\text{max}\{|U|,|B|\}|)^{3}+\mathcal{O}(|\text{max}\{|U|,|N|\}|)^{3}+\mathcal{O}(U\times N) $& $-$&$\frac{\log\left(\frac{M_2}{t^0\varrho}\right)}{\log(\vartheta)}$& $\mathcal{O}(U\times C\times B)$\\
\hline
	\end{tabular}
\end{table}
\begin{table}
\centering
\caption{Complexity order of the proposed solution for caching and delivery phases in NC-OMA.}
\label{NC-OMA}
\begin{tabular}{|L{1cm}|L{2.7cm}|L{1.1cm}|L{1.1cm}|L{1cm}|}
\hline
Phase & User association and subcarrier assignment sub-problem &{Content placement sub-problem}&{Radio resource allocation sub-problem}\\	
\hline
{Caching phase}&$\mathcal{O}(|\text{max}\{|U|,|B|\}|)^{3}+\mathcal{O}(|\text{max}\{|U|,|N|\}|)^{3}+\mathcal{O}(U\times N)$& $\frac{\log\left(\frac{M_4}{t^0\varrho}\right)}{\log(\vartheta)}$&$\frac{\log\left(\frac{M_5}{t^0\varrho}\right)}{\log(\vartheta)}$\\
\hline	
{Delivery phase}&$\mathcal{O}(|\text{max}\{|U|,|B|\}|)^{3}+\mathcal{O}(|\text{max}\{|U|,|N|\}|)^{3}+\mathcal{O}(U\times N) $& $-$&$\frac{\log\left(\frac{M_6}{t^0\varrho}\right)}{\log(\vartheta)}$\\
\hline
\end{tabular}
\end{table}

{
	\section{Numerical Evaluation }\label{Numerical results}
{In this section, the performance of the proposed network is evaluated under different values of the network parameters for
cooperative and non-cooperative systems and also for different multiple access  technologies. We adopt the Monte Carlo method with $ 1,000 $ runs to compute the averaged results. Also, the simulation code was written in MATLAB 2019 and was executed with Core i7 CPU and 8 GB RAM.
\vspace{-0.5em}\subsection{Simulation Environment}
The MBS is considered to be positioned in the center of a circular area with
radius 500 m and $4$ SBSs \cite{7510703} with the circular radius $20$ m are uniformly distributed in the coverage area of MBS. The total
number of users in the network is 50 and all users are distributed uniformly in the coverage of BSs. Moreover, $W^{\text{tot}}$ is set to $20$ MHz based on the LTE \cite{parkvall2009evolution} with the total number of 64 subcarriers.
In this regard, $W $ is set to 312.5 {kHz} \cite{rezvani2019fairness}. $h_{b,u}^{n}(t)=\xi_{b,u}^{n}\cdot(d_{b,u})^{-\kappa}$ where $\kappa$ indicates the path loss exponent which is set to 3 \cite{8786250}, $\xi_{b,u}^{n}$ indicates the Rayleigh fading, and $d_{b,u}$ demonstrates the distance between user $u$ and BS $b$. Moreover, $\sigma^2$ is the power of the AWGN noise and is set to $-174~~\text{dBm}$. We consider the number of contents, i.e., $C=1000$ \cite{7530876, rezvani2019fairness}. The content's popularity is distributed by Zipf where $\alpha= 0.65$. The cache size for MBS and SBSs are assumed to be 10
and 3 percentage of total size of contents, respectively. Also, $P_{0}^{\text{max}}$ and $P_{b}^{\text{max}}$ are considered
as 40 Watts and 2 Watts, respectively \cite{7530876}.
All the parameter values are summarized in Table \ref{table-1}.
}
\begin{table}
\centering
\caption{Simulation setup parameters \cite{7530876, 8611393, 8786250}}
\label{table-1}
\begin{tabular}{|L{1.5cm}|L{2cm}|L{1.5cm}|L{1.8cm}|}
\hline
\cellcolor{yellow}{\textbf{Parameters}}&\cellcolor{yellow}{ \textbf{Values}}&\cellcolor{yellow}{\textbf{Parameters}}& \cellcolor{yellow}{\textbf{Values}} \\
\hline
{MBS radius, SBS radius}&{$500~\text{m},20~\text{m}$}&{$B$} &{$4$}\\
\hline
{$N$}&{ $64$}&{$C$}&{ $1000$}\\
\hline
{$\alpha$ }& {$0.54$}	& {	$\sigma^2$ }& {$-174~\text{dBm}$}\\
\hline
{$C_{\text{BW}}$} & {$3~\text{\$/KHz}$}&{$C_{\text{Power}}$ }& {$5~\text{\$/mWatts}$}\\
\hline
{$C_{\text{FH}}$} &{ $7~\text{\$/KHz}$}&{	$C_{\text{BH}}$} &{ $20\,\text{\$/bps}$}\\
\hline
{$s_c$ distribution parameters} & {$\mu_c=0.5, \sigma_c^{2}=1.5$}&{$\text{M}_{0},\text{M}_b$} &{ $10\%, 3\%$}\\
\hline
{$R^{\text{max,\,FH}}_{i,b}$}&{ $2500~\text{Mbps}$}&{	$R^{\text{max,BH}}_{b}$} & {$\text{Unlimited}$}\\
\hline
{$P_{b}^{\text{max}}$ }&{ $5~\text{Watts}$}&{$P_{0}^{\text{max}}$} & {$40~\text{Watts}$}\\
\hline
{$P_{b}^{\text{mask}}$ }&{ $0.5$~$\text{Watts}$}&{	$l_{b}^{\text{max}}$ }&{ $2$}\\
\hline
{$T$ }&{$300~\mu\text{sec}$}&{$U$}&{$40$}\\
\hline
{$ P_{0}^{\text{Hardware}} $ }&{$5~\text{Watts}$}& $P_{b}^{\text{Hardware}}$&$1~\text{Watts}$\\
\hline
\end{tabular}\label{simval}
\end{table}
\vspace{-0.75em}\subsection{Results Discussions}
{The results of the simulations are shown in Figures  \ref{User_Var}-\ref{SBS}. In what follows, we discuss these results under different caching policies (e.g., ergodic caching) and delivery policies (e.g., cooperative caching) with the advanced access technologies, i.e., OMA and NOMA.
}
\vspace{-0.5em}\subsubsection{The Effect of Number of Users}
Here, we discuss the effect of a number of users on caching and delivery policies cost in the sequel as follows: 
\begin{figure}[t]
	\centering		
	\subfigure[]{ \includegraphics[width=.38\textwidth]{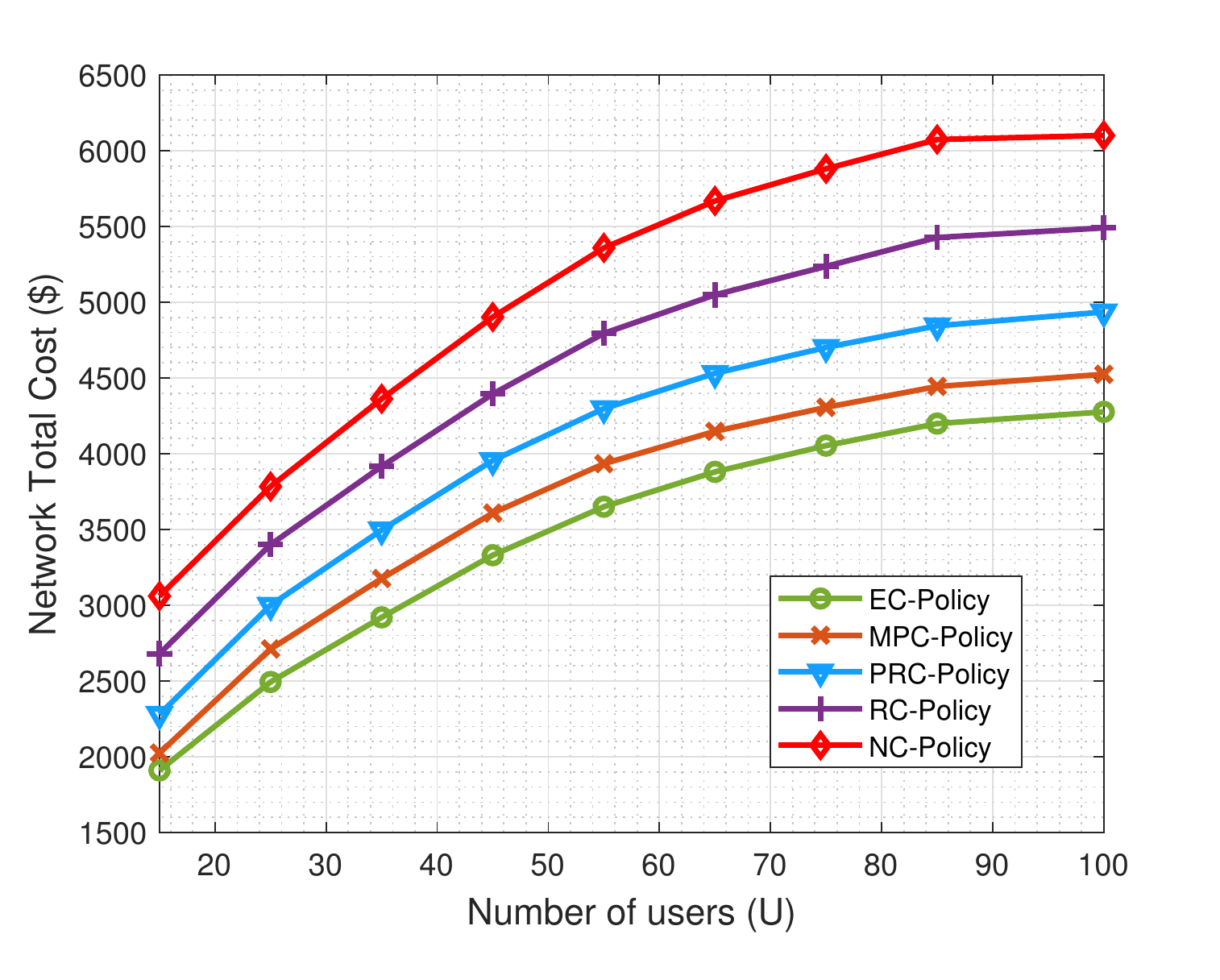}
		\label{UE_Cach}
	}
	\subfigure[]{\includegraphics[width=.38\textwidth]{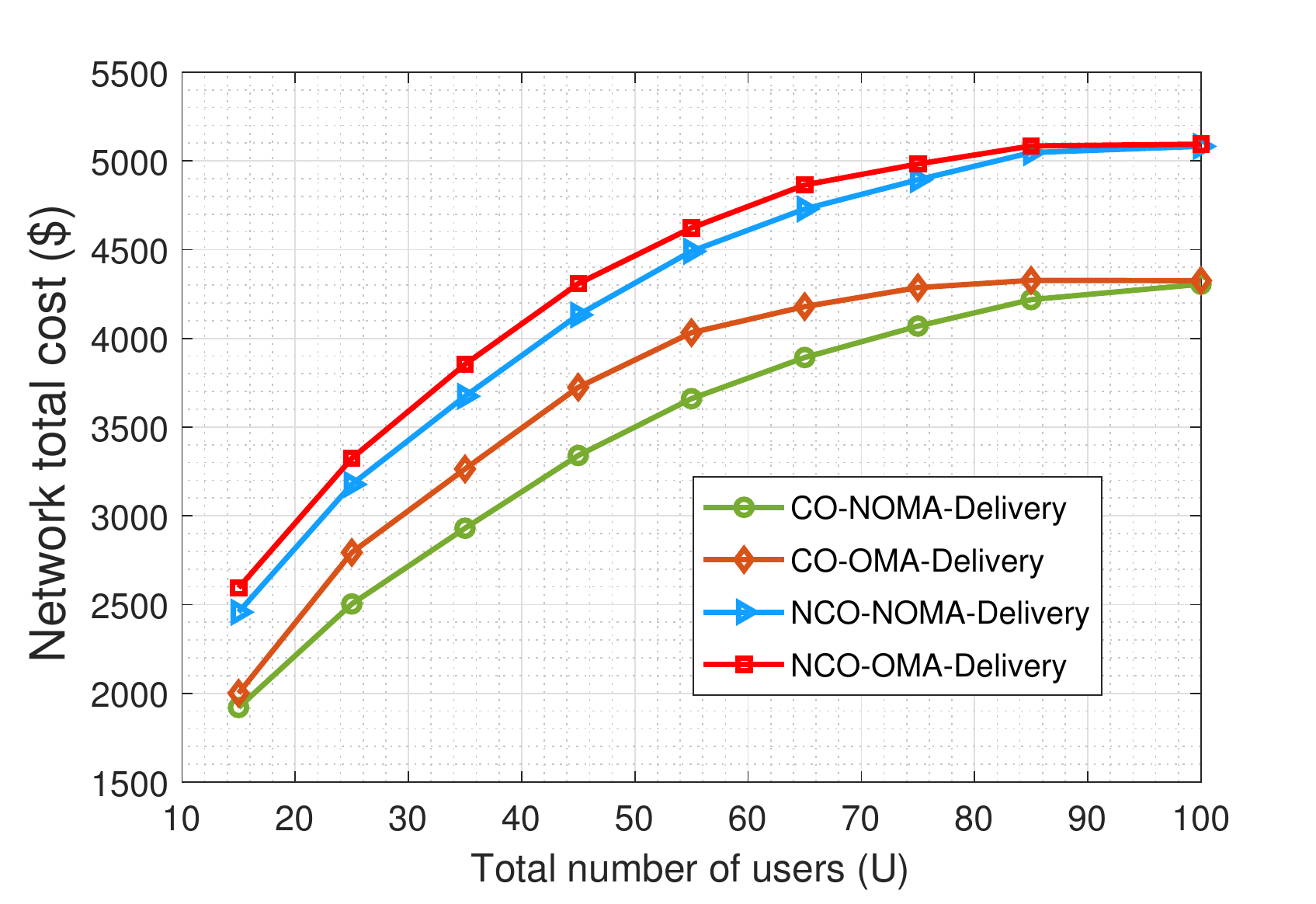}
		\label{UE_Del}
	}
	\subfigure[]{\includegraphics[width=.38\textwidth]{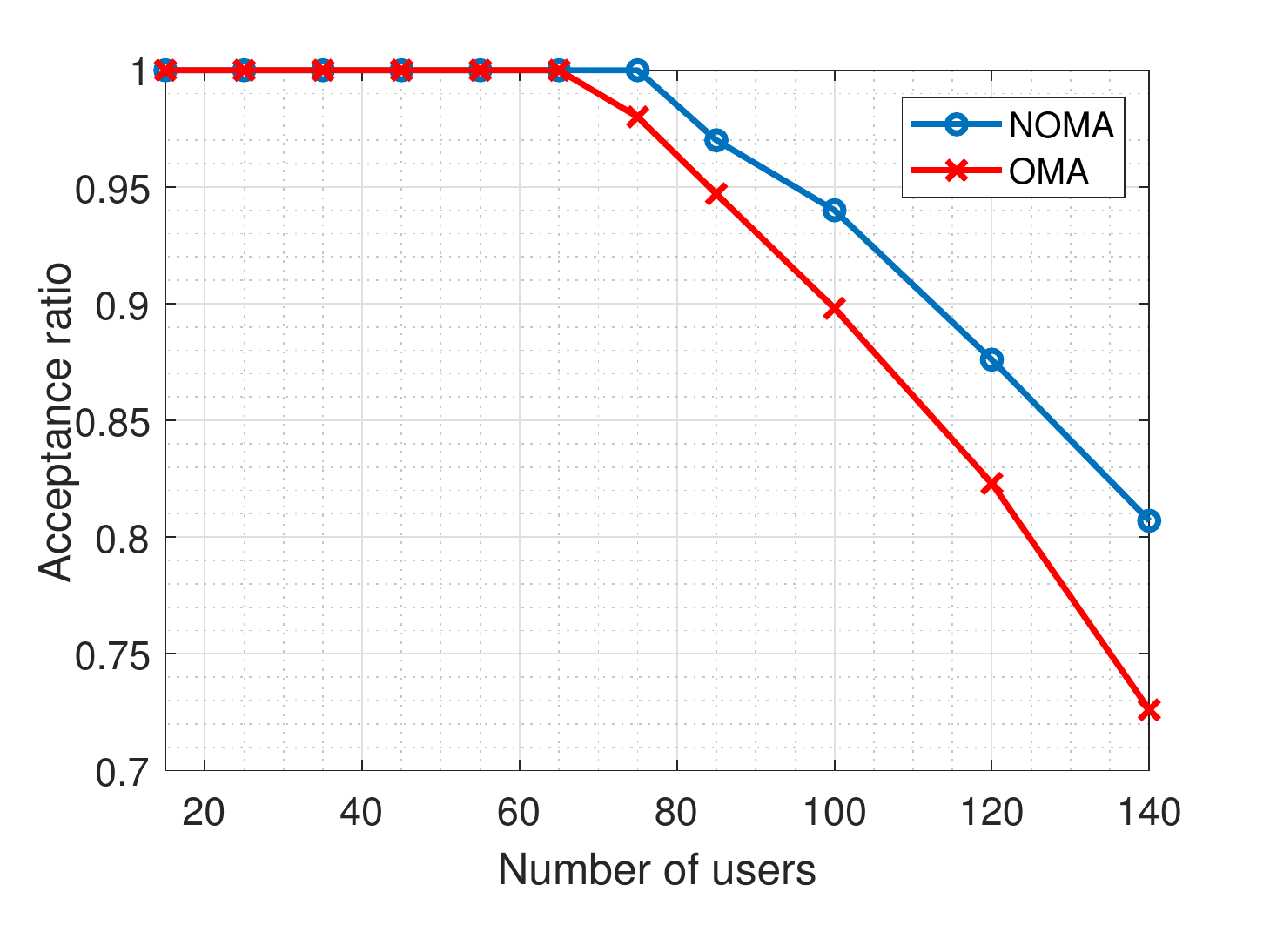}
		\label{AR_U}
	}
	\caption{ The total network cost (a \& b) and acceptance ratio (c) versus the total number of users in the network. 
	}
	\label{User_Var}
\end{figure}
\paragraph{Caching Polices}
{Before discussing on the result of Fig. \ref{User_Var}, we give a brief description on various caching policies for the network. 
No caching (NC) policy {means} there is no cache
storage in all of the BSs \cite{park2016joint}. In this case, all contents need to be transmitted from the remote CP. Random caching (RC) policy caches the contents randomly with uniform distribution. Random caching policy limitation is the BSs cache storage which means that BSs cache contents till their storage is full. Another caching policy is popular random caching (PRC) which is similar to random caching but the caching probability for each content has linear relation with the square root of requesting probability \cite{chen2016probabilistic}. The next
policy is the most popular caching (MPC) policy which is related to the most requested contents \cite{park2016joint}. In this policy, the most
requested contents are cached in the BSs' storages. Also, in this policy, the most popular contents are cached until the BSs' storage is full. The last policy is our proposed ergodic caching policy where the cached contents are determined by the solution of our formulated ergodic resource allocation problem.  
}
{
	\\\indent The delivery cost of the  different caching policies is compared in Fig. \ref{UE_Cach} versus different number of users. It
demonstrates that when the total number of users in the network goes up, the cost of the network increases too. It is
reasonable since the increase of the number of users makes the network to use more resources, thus, the cost of the network goes up.
Also, this figure declares that the worst policy for the network is NC policy. As it is said before,
NC means there is no storages in BSs to cache the contents, hence, all the requested contents need to be fetched from the
backhaul and then transmitted to the users by BSs. 
policies. 
In RC policy, contents are cached in the BSs with the uniform probability. Obviously, there is an expectancy for some
requested contents to be fetched in RAN, either by user's own BS or by the BSs in the neighborhood. Moreover, PRC shows
better performance in the objective compared to RC policy. MPC policy just caches the most popular  contents without
considering the channel condition, and hence it shows lower performance compared to our caching policy. As it is
shown the most effective caching policy for cost minimization is our proposed ergodic caching.
{Our proposed caching policy has much
 	better performance  with the considerable  gain 
 	compared to  MPC and NC policies on average for the number of UEs in the range of $ 15-100 $.
}
}
\paragraph{Delivery Policies}
{We compare the non-cooperative network with PD-NOMA (NC-NOMA) technology, NC-OMA, cooperative policy with PD-NOMA 
(CO-NOMA)  and cooperative network with OMA (CO-OMA) technology with each other in Fig. \ref{UE_Del}. As this figure
expresses, the NC-OMA technology has lower cost compared to the CO-NOMA. It determines that the cooperative delivery case has more
effect on the network cost minimization than PD-NOMA. Moreover, this figure shows our proposed caching policy leads to the most cost reduction. Cooperative delivery policy forces the network to fetch more contents in RAN, thus, less number of requested contents need to be fetched from the backhaul links  which considerably decreases the cost value of the network. Besides, PD-NOMA  drops the cost value by supporting more than one user in each subcarrier, which leads to less spectrum consumption. It is shown that the CO-NOMA has better performance with 27.8571\% network cost minimization compared to NC-OMA.
\\\indent Moreover, from the figure, the costs of OMA and NOMA for both non-cooperative and cooperative delivery policies are approximately close together for number of users more than 85. 
 Actually, by increasing the number of users, utilizing  OMA causes more rejection of users compared to NOMA (See Fig. \ref{AR_U}). In fact, this cost is obtained for a different number of accepted users for OMA and NOMA where NOMA serves more users with the same cost as OMA.  It refers to the wireless capacity gain of NOMA.  
The summary of the achieved results in terms of the cost  for CO-NOMA and NCO-OMA  with different parameters is shown in Table \ref{Gain_Com_T}. 
To obtain a fair statement on the achieved gain, the average on these  
values is taken which  reveals considerable improvement compared to the baseline NCO-OMA.
}
\begin{table*}[!ht]
	\centering
	\tiny
	\caption{Network cost comparison between different delivery polices: cooperative NOMA and non-cooperative OMA}
	\label{Gain_Com_T}
	\begin{tabular}{|l||l|l|l|l|l|l|l|l|l|l|l|l|l|l|l|l|l|l|}
		\hline
		& \multicolumn{7}{c|}{{ Number of Users}} & \multicolumn{6}{l|}{SBSs Cache Size (\%)} & \multicolumn{5}{c|}{Number of SBSs} \\ \cline{2-19} 
		\multirow{-2}{*}{\begin{tabular}[c]{@{}c@{}}Scenario(s)$\rightarrow$ \\ Delivery policy cost value $ \downarrow $\end{tabular}} & 15  & \multicolumn{1}{c|}{25}  & 35  & 45  & 65  & 85  & 100 & 0        & 3        & 6        & 10        & 15        & 20       & 2         & 3        & 4        & 6        & 8        \\ \hline\hline
		CO-NOMA cost value (I)                                                                                           & 1921   & 2521                         & 2973    & 3421    &3971     &4234   &   4305  &     5667     &   4182         &3604         & 2909              &     2604      &  2165        &  5002       &        4694  &        3972  &         3700 & 3550         \\ \hline\hline
		NCO-OMA cost value  (II)                                                                                          & 2594    &  3494                        &   3884  &4503     & 4867  &  5094   &5094     &    5782     & 4867         &   4486       & 3991          &      3687     &  3366        & 5672          &    5341      &        5037  &        4810  &  4653        \\ \hline\hline
		\begin{tabular}[c]{@{}l@{}}(I) over (II) Gain ( \%)  \end{tabular}                                                                                        & 36    &   39                       &   30  &   31  &  23   & 21    &19     & 2         &  16        & 25         &    38       &   42        &   56       &   14        & 14.3         & 27         &  30        &    22  
\\ \hline
	\end{tabular}
\end{table*}
\vspace{-0.5em}
\subsubsection{The SBSs Number Effect}
{The network cost reduction dependency to the number of SBSs is shown in Fig. \ref{FBS}. As it is shown, the growth of the SBSs' number can reduce the network cost substantially. It is obvious that increasing the number of SBSs leads to more caching storage in the network. By increasing the caching storage capacity in the edge, more contents can be cached in the edge of the network, hence, more users can fetch their contents by their own BSs or other BSs in neighbor. In this regard, the usage of costly  backhaul links becomes less frequent, thus, the cost of the network drops. Also, Fig. \ref{FBS} demonstrates four scenarios for the network, which are: 1. CO-NOMA, 2. CO-OMA, 3. NCO-NOMA, and 4. NC-OMA. Like other figures, the worst scenario is NC-OMA because in this case, each subcarrier is used for one user which imposes more power consumption and also, the lack of cooperation in this scenario makes the BSs to fetch the requested contents from backhaul links which lead the network to pay more cost. In contrast with the mentioned scenario, the best performance in cost minimization is CO-NOMA with the greater gain of 21.42\% on average (See Table \ref{Gain_Com_T}) than that of  NC-OMA.
}
\begin{figure}[t]
	\centering
	\includegraphics[width=.37\textwidth]{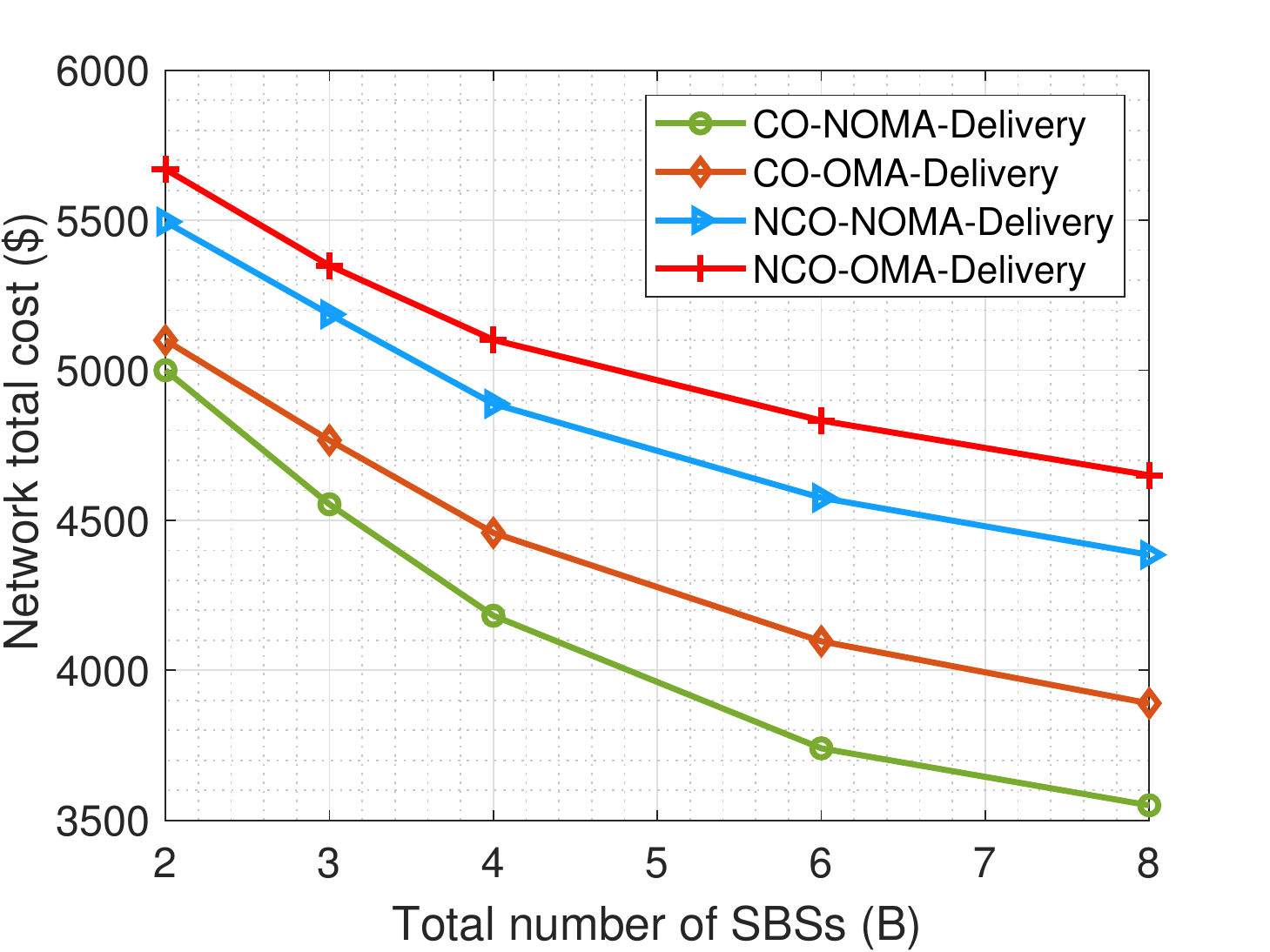}
	\caption{ The total network cost versus the number of SBSs.}
	\label{FBS}
\end{figure}
\begin{figure}[t]
	\centering
	\includegraphics[width=.38\textwidth]{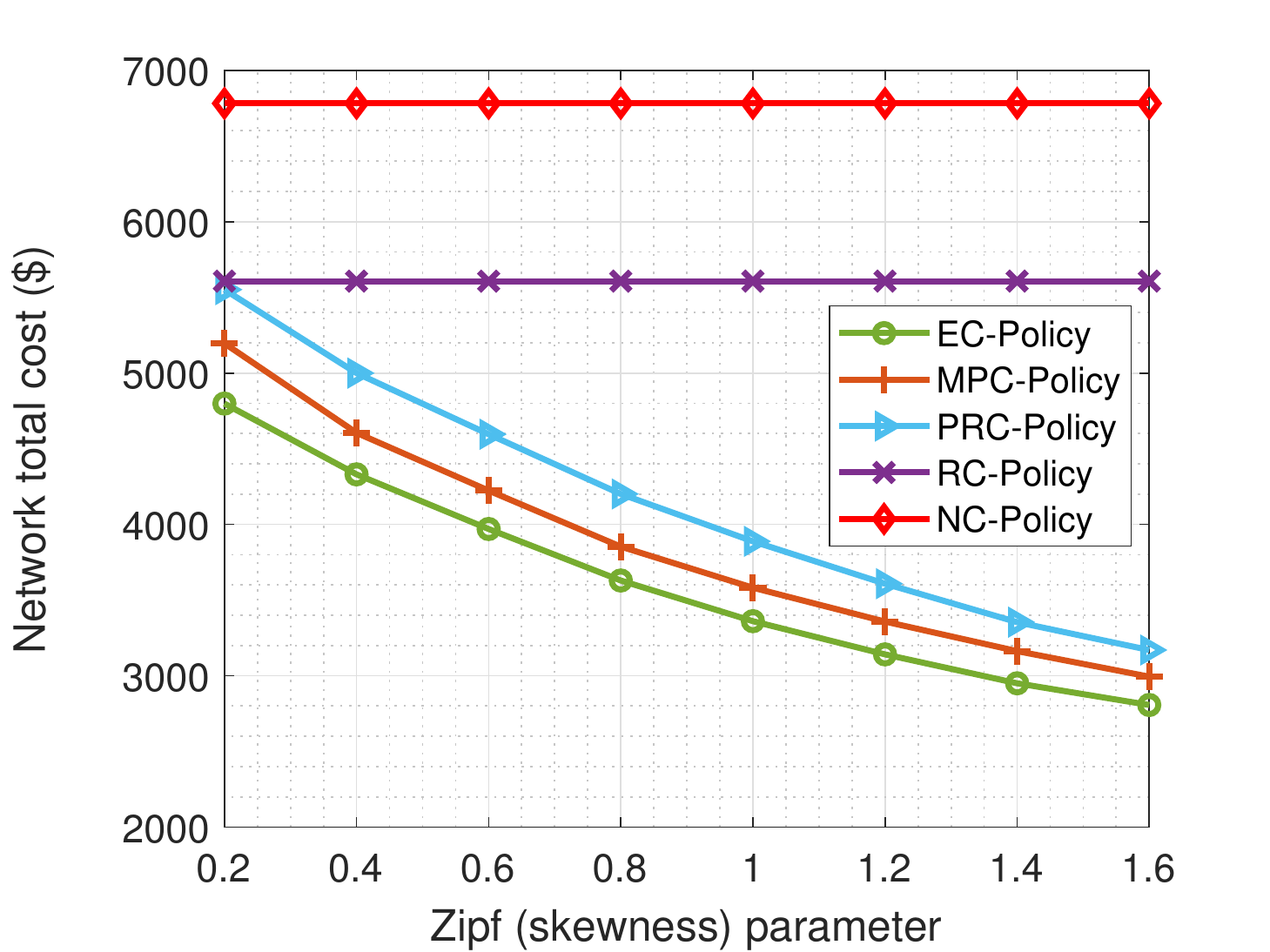}
	\caption{ Total network cost versus Zipf parameters.}
	\label{ZIPf}
\end{figure}
\begin{figure}[t]
\centering
\includegraphics[width=.37\textwidth]{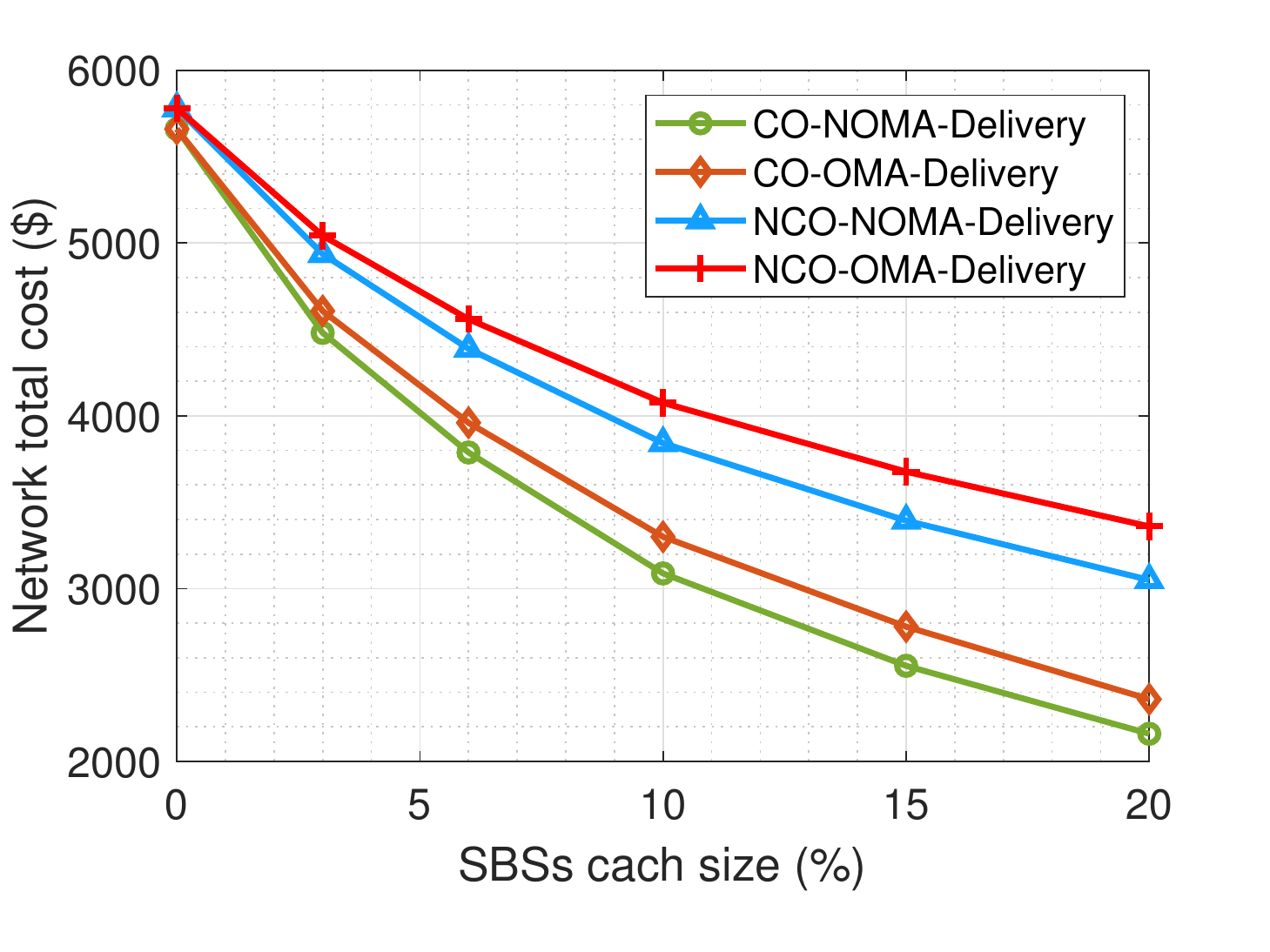}
\caption{ Total network cost versus different SBSs cache size.}
\label{SBS}
\end{figure}
\vspace{-0.75em}\subsubsection{{The Effect of Zipf Parameter}}
{In Fig. \ref{ZIPf}, we investigate the effect of $\alpha$ in our proposed network. $\alpha$ tunes the skewness of PDI and is known as Zipf parameter. When $\alpha$ {becomes} large, the network cost decreases. {This} caused by the increasing probability of requesting identical contents. In this case, the probability of content provisioning from the fronthaul links and even by the associated BS rises up, hence, the network cost drops down.
In NC policy, BSs do not have the ability of caching contents, hence, the Zipf parameter does not have any effect in this case. Also, 
 the RC policy is not affected by the Zipf parameter, since all contents have the same probability (i.e., uniform distribution) independent of the value of $ \alpha $. Moreover, in this figure,  the best caching policy is
ergodic caching polices. Moreover, compared to the NC and RC policies, the ergodic caching gains are considerable (97.28\% and 63.12\%) on average, respectively.
}
\subsubsection{{The Effect of SBSs' Storage Capacity}}
As shown in Fig. \ref{SBS}, if the capacity of SBSs' storage increases, the network cost drops. The reason for network cost decrease is that  there is more place to cache more contents in each the SBSs' storage. Regard to this, the probability of serving the requested contents by the edge of network increases, thus, the network cost drop down.
\\\indent
{NC-OMA network achieves the worst performance which is} based on the non-cooperativeness and OMA technology. 
In CO-NOMA and CO-OMA scenarios, the probability of serving requested contents from the edge is higher than the NCO-NOMA and NC-OMA based on the cooperativeness of BSs. 
Furthermore,  more space
in SBSs' storage provides the CO-NOMA and CO-OMA networks to serve more contents from the edge compared to the NCO-NOMA and NC-OMA networks. Also, PD-NOMA technology is another promising way to decrease the network cost by assigning a subcarrier to more than one user. Thus, the CO-NOMA networks have the most network cost reduction, and have better performance about 26.382\% on average (See Table \ref{Gain_Com_T}) compare to the NC-OMA networks.
\vspace{-0.75em}\subsubsection{Optimality Gap}\label{Optimality_Gap}
Here, we intend to evaluate the optimality gap of the original problem by adopting the exhaustive search method. As known, the computational complexity of the exhaustive search is prohibitive
which exponentially grows with the size and number of parameters, and hence, we only apply this method to a small-scale network. The obtained results in terms of the global optimality gap are presented in Fig. \ref{Optimality_Gap_S}. As seen, Fig. \ref{Optimality_Gap_S} shows the values of the objective function (network cost) for three solution algorithms: 
1) \textit{Circle line/green color:}  the objective function  with the optimal solution of the original problem found by brute force approach;
2) \textit{Cross marked/blue color:} the objective function with the iterative solution, but in that way  the content delivery sub-problem is solved via its optimal solution \eqref{sub-problem1};
3) \textit{Diamond line/red color:} the objective function with the iterative solution solving the content delivery sub-problem \eqref{sub-problem1} by the proposed heuristic Alg. \ref{heuristic}. As can be seen from the figure, 
the global optimality gap of the original problem is small in overall, and 
the  heuristic algorithm of content delivery sub-problem is near optimal in all the considered cases.		
	\begin{figure}[!ht]
	\centering
	\includegraphics[width=0.4\textwidth]{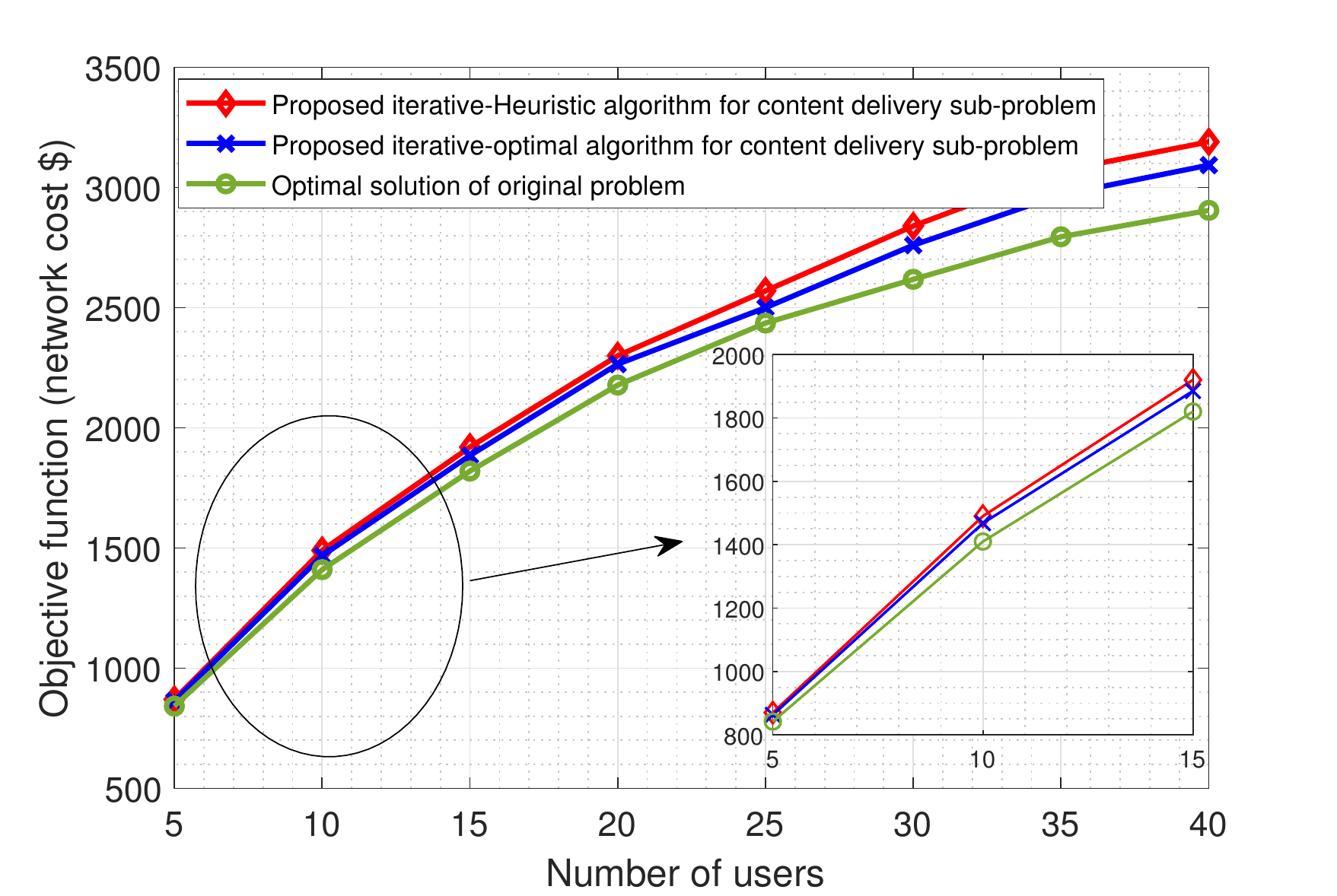}
	\vspace{-1.25em}
	\caption{The global optimality gap analysis for different algorithms.} 
	\label{Optimality_Gap_S}
\end{figure}
\vspace{-1.5em}
\section{Conclusion}\label{concolusion}
{In this work, we proposed a cooperative caching framework for HetNets which {adopts} PD-NOMA. We formulated a novel cost function and proposed an optimization problem to minimize the network cost. Also, in our scheme, we considered two phases: caching phase and delivery phase. In the first phase, the network caches the contents based on their PDI jointly with the ergodic resource allocation. In the second phase, we deliver all the requested contents to the users. Moreover, to solve each phase's MINLP, the ASM is applied to handle the non-convexity of the optimization problem. We also employed the Hungarian algorithm, {SCA}, and a novel heuristic algorithm to solve each phase's optimization problem. 
{Simulation results showed that the devised schemes improves the cost reduction compared to other caching policies and OMA. The proposed scheme can considerably reduce  (See Table \ref{Gain_Com_T}) the network cost compared to NC-OMA  {which verifies the effectiveness of our proposed scheme}.
}
}}
\appendices
\section*{Appendix A}
\paragraph*{\textbf{\textit{The Initial Point of the Caching Phase}}}
To utilize ASM in the caching phase, we need to provide a feasible initial point. To this end, we use the elasticization method and introduce $A\geq 0$ as an elastic variable [66]. In this method, $A$ is introduced to all the constraints and we aim to minimize the margin:
\begin{subequations}
	\label{initial}
	\begin{align}
	&
	\minimize_{\boldsymbol{p, \tau, \rho, A}}A \nonumber\\
	&
	\text{s.t. }\nonumber\\
	&
	\forall b\in\mathcal{B}, c\in\mathcal{C},\nonumber\\
	&
	\label{cpi7}
	A \geq
	-\Bbb{E}_{\bold{H}(t)}\{\frac{h_{b,u}^{n}(t)}{I_{b,u}^{n,\text{Intra}}(t)+I_{b,u}^{n,\text{Inter}}(t)+\sigma^2}\}+\\
	&
	\Bbb{E}_{\bold{H}(t)}\{\frac{h_{b,u'}^{n}(t)}{I_{b,u'}^{n,\text{Intra}}(t)+I_{b,u'}^{n,\text{Inter}}(t)+\sigma^2}\}, \,\ \nonumber\\
	&
	\forall b\in\mathcal{B}, u, u' \in \mathcal{U},u'\neq u ,h_{b,u}^{n}(t) \geq h_{b,u'}^{n}(t), n\in\mathcal{N},\nonumber\\
	&
	\label{cpi10}
	\sum_{c=1}^{C}d_{c}\cdot\rho_{b.c}\cdot s_{c} - \sum_{u=1}^{U}\sum_{n=1}^{N}T\Bbb{E}_h\{r^{n}_{b,u}(t)\})
	\leq A, \,\ \forall b\in\mathcal{B}, \\
	&
	A\geq 0,\\
	&
	(10a), (10b), (10d), (10e), (10g)-(10i):.\nonumber
	\end{align}
\end{subequations}
{
	If $A=0$, the feasible initial point is achieved, otherwise there is an essential need to have an admission control {to reduce the number of admitted users} which is performed based on our previous work [55].
} Moreover, (48) is MINLP and (48b) is non-convex. Hence, we apply ASM to propose an algorithm for the initial content placement, initial subcarrier assignment, and initial resource allocation values. The steps of the algorithm are similar to ones in the algorithms in Subsections 3.2.1, 3.2.2, and 3.2.3, respectively.
\section*{Appendix B}
\paragraph*{\textbf{\textit{The Initial Point of the Delivery Phase}}}
We need to provide the feasible initial point in the delivery phase. Regard to this, we calculate the initial point, based on satisfying the corresponding constraints of (37). Furthermore, we introduce $M$ as an elastic variable to utilize the elasticizing approach. In this case, all the constraints which make (37) infeasible are changed as follows. Based on this method, the elastic variable $M\geq 0$ and is introduced into the optimization problem. The following problem defines the elastic optimization problem:
\begin{subequations}
	\label{initiald}
	\begin{align}
	&
	\minimize_{\boldsymbol{r^{\text{BH}}, r^{\text{FH}}, p, \tau, \rho, M}}M \nonumber\\
	&
	\text{s.t. }\nonumber\\
	&
	\label{cpii1}
	\min _{u , \delta_{u,c}=1}\{\sum_{n=1}^{N}r_{b,u}^{n}\}y_{i,b,c}- y_{i,b,c}r_{i,b,c}^{\text{FH}}\leq M, \,\ \forall i, b\in\mathcal{B}, i\neq b,\\
	&
	c\in\mathcal{C},\nonumber\\
	&
	\label{cpii2}
	\min _{u , \delta_{u,c}=1}\{\sum_{n=1}^{N}r_{b,u}^{n}\}z_{b,c}-z_{b,c}r_{b,c}^{\text{BH}}\leq M, \,\  \forall b\in\mathcal{B}, c\in\mathcal{C},\\
	&
	\label{cpii7}
	M \geq\\
	&
	-\frac{h_{b,u}^{n}}{I_{b,u}^{n,\text{Intra}}+I_{b,u}^{n,\text{Inter}}+\sigma^2}+
	\frac{h_{b,u'}^{n}}{I_{b,u'}^{n,\text{Intra}}+I_{b,u'}^{n,\text{Inter}}+\sigma^2}, \,\ \nonumber\\
	&
	\forall b\in\mathcal{B}, u, u' \in \mathcal{U},u'\neq u ,h_{b,u}^{n} \geq h_{b,u'}^{n}, n\in\mathcal{N},\nonumber\\
	&
	\label{cpii10}
	\sum_{n=1}^{N}\delta_{u,c}\tau_{b,u}^{n}s_c-\delta_{u,c} (\sum_{n=1}^{N}\tau_{b,u}^{n}T)(\sum_{n=1}^{N}r^{n}_{b,u})\leq M, \,\ \forall b\in\mathcal{B},\\
	&
	c\in\mathcal{C}, u\in\mathcal{U},\nonumber \\
	&
	\label{cpii13}
	\sum_{c=1}^{C}r_{i,b,c}^{\text{FH}}- R_{i,b}^{\text{max},\text{FH}}\leq M, \,\ \forall i, b\in\mathcal{B}, i\neq b,\\
	&
	M\geq 0,\\
	&
	(5), (6), (9), (24), (25), (26).\nonumber
	\end{align}
\end{subequations}
{
	If $M=0$, the feasible initial point is achieved, otherwise there is an essential need to have an admission control {to reduce the number of admitted user} which is performed based on our previous work [55].
}
%
%
Moreover, (49) is MINLP and (49a), (49b) and (49d) are non-convex. Hence, we apply ASM. The proposed algorithms for the initial point optimization problem are similar to the applied algorithms in Subsections 3.2.2, 4.2.2 and 4.2.3, respectively.
\section*{Appendix C}
\paragraph*{\textbf{\textit{Convergence for the ASM in the Delivery Phase}}}
Based on the ASM method, after each iteration, the objective function in each sub problem is enhanced and finally it
converges. {Fig. 6} shows the value of the objective function versus the number of iterations for different number of users and the maximum transmission power. From this figure, we can see that after three iterations, the value of the objective
function is fixed. For overall algorithm described in Alg. 1, after applying the first step, obtaining $\boldsymbol{\rho}$ of iteration $s+1$ with given $\boldsymbol{\tau}^{(s)}=\boldsymbol{\tau}^{(s)}$,  $\boldsymbol{r}^{\text{FH}}=\boldsymbol{r}^{\text{FH},(s)}$, $\boldsymbol{r}^{\text{BH}}=\boldsymbol{r}^{\text{BH},(s)}$ and $\boldsymbol{p}=\boldsymbol{p}^{(s)}$, we have
\begin{align}
&
\phi(\boldsymbol{r}^{\text{BH},(s)}, \boldsymbol{r}^{\text{FH},(s)}, \boldsymbol{p}^{(s)}, \boldsymbol{\rho}^{(s)}, \boldsymbol{\tau}^{(s)}
)\geq\nonumber\\
&
\phi(\boldsymbol{r}^{\text{BH},(s)}, \boldsymbol{r}^{\text{FH},(s)}, \boldsymbol{p}^{(s)}, \boldsymbol{\rho}^{(s+1)},
\boldsymbol{\tau}^{(s)}).
\end{align}
Moreover, after applying the second step, obtaining $\boldsymbol{\tau}^{(s)}$ of iteration $ks+1$ with given
$\boldsymbol{\rho}=\boldsymbol{\rho}^{(s+1)}$, $\boldsymbol{r}^{\text{FH}}=\boldsymbol{r}^{\text{FH},(s)}$,
$\boldsymbol{r}^{\text{BH}}=\boldsymbol{r}^{\text{BH},(s)}$ and $\boldsymbol{p}=\boldsymbol{p}^{(s)}$ we have
\begin{align}
&
\phi(\boldsymbol{r}^{\text{BH},(s)}, \boldsymbol{r}^{\text{FH},(s)}, \boldsymbol{p}^{(s)}, \boldsymbol{\rho}^{(s+1)}, \boldsymbol{\tau}^{(s)},
)\geq\nonumber\\
&
\phi(\boldsymbol{r}^{\text{BH},(s)}, \boldsymbol{r}^{\text{FH},(s)}, \boldsymbol{p}^{(s)}, \boldsymbol{\rho}^{(s+1)},
\boldsymbol{\tau}^{(s+1)}).
\end{align}
Furthermore, after applying the third step, obtaining $\boldsymbol{r}^{\text{FH}}$, $\boldsymbol{r}^{\text{BH}}$ and, $\boldsymbol{p}$ of iteration $s+1$ with given
$\boldsymbol{\rho}^{(s+1)}$, $\boldsymbol{\tau}^{(s+1)}$, we have
\begin{align}
&
\phi(\boldsymbol{r}^{\text{BH},(s)}, \boldsymbol{r}^{\text{FH},(s)}, \boldsymbol{p}^{(s)}, \boldsymbol{\rho}^{(s+1)}, \boldsymbol{\tau}^{(s+1)})
\geq\nonumber \\
&
\phi(\boldsymbol{r}^{\text{BH},(s+1)}, \boldsymbol{r}^{\text{FH},(s+1)}, \boldsymbol{p}^{(s+1)}, \boldsymbol{\rho}^{(s+1)}, \boldsymbol{\tau}^{(s+1)}),
\end{align}
due to the fact that for a given feasible amount for the caching phase, the optimization problem improves the objective
function after each iteration. Finally, we have a sub-optimal solution as the output of the algorithm.
\section*{Appendix D}
\paragraph*{\textbf{\textit{Convergence of SCA with DC Approximation}}}
The SCA approach with DC approximation generates a sequence of improved feasible solutions which converges to a sub-optimal. For notation simplicity, we omit parameter $ t $.
$g_{b,u}^{n,(k)}$ is well approximated to its first order approximation (17). Then, function (14) is approximated with the concave function
(19). Since $g_{b,u}^{n,(k)}$ is a concave function, we have
$g_{b,u}^{n,(k)}\leq g_{b,u}^{n,(k-1)}+\bigtriangledown g_{b,u}^{n,(k-1)}(\boldsymbol{p}^{(k)}-\boldsymbol{p}^{(k-1)}).$
As well, after iteration \textit{k}, we note that:
\begin{align}\label{proof2}
&
\sum_{n=1}^{N}\delta_{u,c}^{(k)}\tau_{b,u}^{n,(k)}s_c\leq\delta_{u,c}^{(k)} (\sum_{n=1}^{N}\tau_{b,u}^{n}T)\big(\sum_{n=1}^{N}\big(f_{b,u}^{n,(k)}-\{g_{b,u}^{n,(k-1)}+\nonumber\\
&\bigtriangledown g_{b,u}^{n,(k-1)}(\boldsymbol{p}^{(k)}-\boldsymbol{p}^{(k-1)})\}\big)\big).
\end{align}
Moreover, $f_{b,u}^{n}$ is well approximated to its first order approximation (37). Since $f_{b,u}^{n,(k)}(\boldsymbol{p})$ is a concave function, we have
$f_{b,u}^{n,(k)}\leq f_{b,u}^{n,(k-1)}+\bigtriangledown f_{b,u}^{n,(k-1)}(\boldsymbol{p}^{(k)}-\boldsymbol{p}^{(k-1)}).$
As well, after iteration $(k-1)$, we note that
\begin{align}\label{proof4}
&
{\min _{\scriptstyle ~~~u\hfill\atop{\scriptstyle~\delta_{u,c}^{(k)}=1}}}\{\sum_{n=1}^{N}(g_{b,u}^{n,(k-1)}-\{f_{b,u}^{n,(k-1)}+\nonumber\\
&
\bigtriangledown f_{b,u}^{n,(k-1)}(\boldsymbol{p}^{(k)}-\boldsymbol{p}^{(k-1)})\})\}y_{i,b,c}\leq
y_{i,b,c}r_{i,b,c}^{\text{FH}}.
\end{align}
and
\begin{align}
&
{\min _{\scriptstyle ~~~u\hfill\atop{\scriptstyle~\delta_{u,c}=1}}}\{\sum_{n=1}^{N}(g_{b,u}^{n}-\{ f_{b,u}^{n,(k-1)}+\nonumber\\
&
\bigtriangledown f_{b,u}^{n,(k-1)}(\boldsymbol{p}^{(k)}-\boldsymbol{p}^{(k-1)})\})\}z_{b,c}\leq z_{b,c}r_{b,c}^{\text{BH}}.
\end{align}
Since the constraints set is compact and the objective function is improved in each iteration, SCA-DC is converged to a sub-optimal solution.
{Fig. 6(a)} {demonstrates the fast convergence of  the proposed SCA.}
\section*{Appendix E}
\subsection*{\textbf{\textit{Computational Complexity}}}\label{Computational complexity}
The complexity order of the solution for ASM is the sum of all the problems computational complexity function. Each sub-problem computational complexity depends on the number of constraints and the required number of iterations.
In particular, the required number of iterations for each phase depends on the convergence of each sub-problem and the accuracy of the applied algorithm.\\
For subcarrier assignment sub-problems in both the caching phase and the delivery case, the Hungarian algorithm computational complexity is calculated as follows [8]:
\begin{align}
\mathcal{O}(|\text{max}\{|U|,|B|\}|)^{3}+\mathcal{O}(|\text{max}\{|U|,|N|\}|)^{3}.
\end{align}
In the DC approximation, the primary computational complexity originates from solving the problem via CVX. The computational complexity for resource allocation problems in the caching phase is as follows:
\begin{align}
\Lambda_1=\frac{\log(M_1/t^{0}\vartheta)}{\log\zeta},
\end{align}
where $M_1$ is the total number of constraints in (21) which is equal to
{\begin{align}
	&
	M_1=2\cdot [B+U\times B\times N],
	\end{align}
}
where $\zeta$ is used for the accuracy updating of the interior point method (IPM), $0<\varrho\ll1$ is the stopping criterion for IPM and $t^0$ is the initial point for approximating the accuracy of IPM. According to (42), the computational complexity for resource allocation sub-problem in the delivery phase is as follows:
\begin{align}
\Lambda_2=\frac{\log(M_2/t^{0}\vartheta)}{\log\zeta},
\end{align}
{where $M_2$ is the total number of constraints of (42) and is given by (46).}
Also, the computational complexity for the content placement problem is
\begin{align}
\Lambda_3=\frac{\log(M_3/t^{0}\vartheta)}{\log\zeta},
\end{align}
{where $M_3$ is the total number of constraints in (11) which is equal to
	\begin{align}
	\nonumber
	M_3= 2\cdot B+B\times C=\mathcal{O}\left(B\times C\right). 
	\end{align}
}
The computational complexity for the delivery cases sub-problem is in the order of $\mathcal{O}(U\times C\times B)$. All the
computational complexity functions are summarized in Table 3. In the next subsection, we compare our proposed network with the non-cooperative network with OMA (NC-OMA) from the computational complexity perspective.

\bibliographystyle{ieeetr}
\bibliography{citation_Caching_2020-06-10}
\end{document}